\documentclass[twocolumn]{article}
\usepackage{graphicx}
\usepackage{aas_macros}
\usepackage{amssymb}

\def\be{\begin{equation}}
\def\ee{\end{equation}}
\def\bea{\begin{eqnarray}}
\def\eea{\end{eqnarray}}

\def\narrowfig{\columnwidth}
\def\widefig{2\columnwidth}

\def\citeNP#1{\cite{#1}}

\def\lsim{\mathrel{\rlap{\lower4pt\hbox{\hskip1pt$\sim$}}
    \raise1pt\hbox{$<$}}}           
\def\gsim{\mathrel{\rlap{\lower4pt\hbox{\hskip1pt$\sim$}}
    \raise1pt\hbox{$>$}}}           

\def\bi#1{\mathbf{#1}}
\def\trace{{\rm tr\ }}

\begin{document}

\twocolumn[
\begin{@twocolumnfalse}
\title{Prospects for ACT: simulations, power spectrum, and 
non-Gaussian analysis}
\author{Kevin M. Huffenberger, Uro\v s Seljak\\
\textit{Department of Physics, Jadwin Hall, Princeton University, Princeton, 
NJ 08544, USA}}
\maketitle
\begin{abstract}
A new generation of instruments will reveal the microwave sky at high 
resolution.  We focus on one of these, the Atacama Cosmology Telescope, 
which probes scales $1000<l<10000$, where both primary and secondary 
anisotropies are important.  Including lensing, thermal and kinetic 
Sunyaev-Zeldovich (SZ) effects, and extragalactic point sources, 
we simulate the telescope's observations of the CMB in three 
channels, then extract the power spectra of these components in a 
multifrequency analysis.
We present results for various cases, differing in 
assumed knowledge of the 
contaminating point sources.  
We find that both radio and infrared point sources are important, 
but can be effectively 
eliminated from the power spectrum given three (or more) channels and 
a good understanding of their frequency dependence.  However, 
improper treatment of the scatter in the point source frequency 
dependence relation may introduce a large systematic bias.
Even if all thermal SZ and
point source effects are eliminated, the kinetic SZ effect 
remains and corrupts measurements of the primordial slope and amplitude
on small scales. We discuss 
the non-Gaussianity of the one-point probability 
distribution function as a way 
to constrain the kinetic SZ effect, 
and we develop a method for distinguishing this effect from the CMB 
in a 
window where they overlap.  This method provides an independent 
constraint
on the variance of the CMB in that window and is complementary to the 
power spectrum analysis. 
\end{abstract}
\end{@twocolumnfalse}
]
  
  \section{Introduction}
Several survey telescopes 
(SPT\footnote{http://astro.uchicago.edu/spt/}, 
APEX\footnote{http://bolo.berkeley.edu/apexsz/}, 
ACT\footnote{http://www.hep.upenn.edu/act/act.html}) will open 
a poorly probed regime of the microwave background: 
arcminute scales at frequencies up to a few hundred GHz. 
The sensitivities of these instruments will be a few micro-Kelvin. 
These 
efforts will provide a large amount of high-quality data, but face a 
different set of challenges than lower resolution surveys, such 
as {WMAP}\footnote{http://map.gsfc.nasa.gov/}.  For lower resolution 
experiments  the primary anisotropies of the CMB are the principal 
signal.  The major contaminant is the galaxy, whose emission is 
diffuse and drops off rapidly away from the galactic center. 
By contrast, for higher resolution experiments, secondary anisotropies 
dominate at small scales.  Assuming the 
observations are away from the galaxy at high frequencies, the major 
contaminant is extragalactic point source emission.  
For a wide range of interesting scales, point sources contribute 
substantially more power than the instrument noise.

The statistical properties of the primary CMB differ from those of 
these signals. Primary anisotropies, from the surface of last 
scattering, dominate the CMB on degree scales.  
The fluctuations are Gaussian, so the power 
spectrum sufficiently describes the anisotropy.  
Secondary anisotropies, such as lensing and the Sunyaev-Zeldovich 
(SZ) effect, 
depend on the late-time, nonlinear evolution of the universe. Their 
imprints need not be Gaussian.  

In this work, we investigate the prospects of these experiments to 
extract the power spectrum and other statistical 
information for the components that contribute
at these angular scales and frequencies. We are particularly 
interested
in determining the primordial power spectrum from the 
primary CMB. Such a determination would enable one to determine the 
amplitude and slope of the fluctuations on small scales and, in 
combination 
with large scale observations, would allow one to place powerful 
constraints 
on models of structure formation. 
We model our investigations on
 the Atacama Cosmology Telescope (ACT), a 6 meter off-axis telescope 
to be placed in the Atacama desert in the mountains of northern Chile 
\cite{2003NewAR..47..939K}.  ACT's bolometer array will measure in 
three bands from 145--265 GHz, with beams sized 1--2 arcminutes.  
(See Table \ref{tab:actspecs}.)  ACT will survey about 100 square 
degrees of the sky with high signal to noise.  
In section \ref{sec:sims}, we simulate observations based on the 
specifications of ACT.  

The first problem in extracting the primary CMB information 
from observations is separating
 the components on the sky with distinct frequency dependences.  
Although many signals  at these 
scales are not Gaussian, the power 
spectrum is still the most important of the statistics, 
and in section \ref{sec:2point} we extract 
the power spectrum with a multifrequency analysis.  The power 
spectrum analysis requires 
knowledge of the frequency dependence of the signals 
and of the scatter in that relation.  Our knowledge is presently 
incomplete, and any error biases the estimated
power spectra.  Our discussion explores the impact of estimating some 
of this frequency information from the data itself.

The second problem in extracting the primary CMB information 
 is that one of the secondary anisotropies, the kinetic 
Sunyaev-Zeldovich (kSZ) effect (and its analogs like the 
Ostriker-Vishniac effect 
and patchy reionization effect), has the same frequency dependence 
as the primary CMB.  This complicates their separation, requiring 
an explicit template for the kSZ power spectrum. Again our knowledge 
is incomplete, and any error biases the primary CMB power spectrum.  
We address in detail the impact of secondary anisotropies on the 
estimation of the amplitude and slope of primordial fluctuations.

Non-Gaussian information can help
distinguish between the components which cannot be separated using 
frequency information.  
It is difficult to make non-Gaussian analysis fully exhaustive, so one must 
choose some
simple statistics. Here
we explore the one-point probability 
distribution function, in section \ref{sec:ng-pdf}.  This can provide 
a consistency check on a kSZ template and can assist in signal 
separation.  
Finally, in section \ref{sec:conclusions} we present our conclusions.

\begin{table*}
\begin{center}
\begin{tabular}{|c|c|c|} \hline
	Band  (GHz) & Beam FWHM (arcmin.) & Sensitivity per pixel 
($\mu$K) \\ \hline
		145       & 1.7			& 2		
	   \\
		210	  & 1.1			& 3.3		
	   \\
		265	  & 0.93		& 4.7		
	   \\ \hline
\end{tabular}
\caption{Frequency channels, beam full-width at half maximum, and 
detector noise (thermodynamic temperature) of mock ACT 
survey.}\label{tab:actspecs}
\end{center}
\end{table*}

\section{Simulations}\label{sec:sims}
\def\n{\hat {\mathbf n}}
\def\bn{{\mathbf n}}

  We simulate the sky as five astrophysical components (CMB, kSZ, SZ, 
and two types of point sources) plus noise.  The CMB is a Gaussian 
random field.  The density field (used for lensing of CMB), 
kSZ, and SZ are produced from 
hydrodynamical simulations \cite{2002ApJ...577..555Z}.  Point sources are 
Poisson distributed 
from source count models.  We add the appropriate detector noise and 
convolve with the telescope beam (assumed Gaussian).

We briefly examine the simulations in general, before addressing each 
signal in turn.  The simulation contains forty fields, each 
$1.19^\circ \times 1.19 ^\circ$, totaling  $\sim 57$ square degrees, 
somewhat smaller than the ACT survey.  The results will therefore be 
appropriate for a smaller, more conservative mock survey.  
Throughout, we have used the following cosmological parameters in a 
flat, $\Lambda$CDM model:  $\Omega_c=0.32$, $\Omega_b=0.05$, 
$\Omega_\Lambda=0.63$, $\sigma_8=1.0$, with $H_0=70$ km/s/Mpc.  
  
\begin{figure}
\begin{center}
\includegraphics[width=\narrowfig]{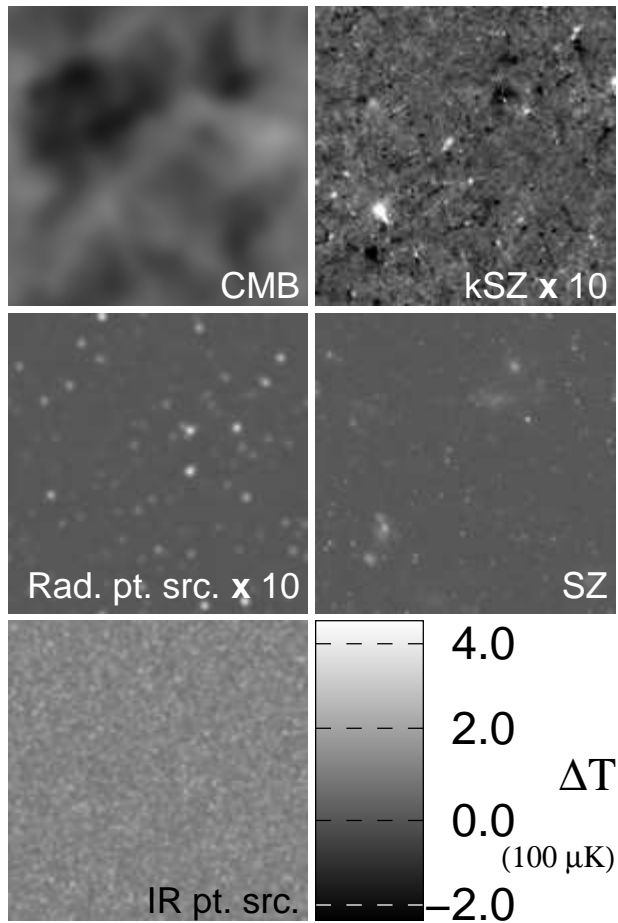}
\end{center}
\caption{These simulated signals contribute to one of forty 
$1.2^\circ \times 1.2 ^\circ$ fields.  All fields are in temperature 
units with a common gray scale, although kSZ and radio source signals 
have been multiplied by 10 for visibility.  SZ is shown at 
$-T_{RJ}=2y$.  The CMB, SZ, and kSZ are not convolved with any beam.  
For visibility, radio and infrared point sources are convolved with 
the beam of the channel in which are most prominent, respectively a 
1.7 arcminute (145 GHz channel) and a 0.9 arcminute (265 GHz) 
full-width at half-maximum Gaussian beam.}\label{fig:signals}
\end{figure}

We express our fields in temperature units, or as a temperature 
perturbation $\Theta(\n) = \Delta T(\n)/T_{\rm CMB}$ for line of 
sight $\n$.   Sample 
temperature perturbation maps of each signal are shown in Figure 
\ref{fig:signals}.  We will refer to each image as we discuss the 
signals in greater detail.  Signals which have a blackbody frequency 
dependence (CMB, kSZ) have the same temperature perturbation at all 
observed frequencies.  For all other signals, the temperature 
perturbation will depend on the observation frequency.

In Fourier space, we use the flat sky approximation, 
\be
\Theta(\n) = \frac{1}{(2\pi)^2}\int d^2{\mathbf l}\  \Theta_{\mathbf 
l} \exp(i \n \cdot {\mathbf l}).
\ee 
The symbol $\mathbf l$ is a two dimensional wavevector.  
The power spectrum of a signal or channel is defined by $\langle 
\left| \Theta_{\mathbf l} \right|^2 \rangle = C_l$.   For each 
signal, figure \ref{fig:crosspower} displays the power spectra and 
cross-correlations between the bands. Below we include a more 
detailed look at these spectra.

ACT will have channels near 145, 217, and 265 GHz, which in 
simulations we take to have delta function frequency responses.  These 
bands are typical of ground based experiments because they lie in 
windows between atmospheric lines.  The band at 217 GHz also takes 
advantage of the thermal SZ ``null,'' described below in section 
\ref{subsec:sz}.  Figure \ref{fig:channels} show a simulated patch of 
sky observed by ACT in these three bands.  We discuss this image 
further below.

In the next section, we discuss the hydrodynamical simulations which 
were the source of several of our signals.  Then we study the signals 
and noise in more detail.  Section \ref{subsec:simcaveats} list 
caveats to our simulations.  Section \ref{subsec:sigsum} summarizes 
the discussion of the simulations.

\subsection{Hydrodynamical simulations}

   Hydrodynamical simulations provide the 
density field for lensing, the SZ and kSZ effects.  These simulations 
are described in more detail elsewhere 
\cite{1998ApJS..115...19P,2001PhRvD..63f3001S,2002ApJ...577..555Z,2003astro.ph..4534Z}.  
A 100 $h^{-1}$ Mpc periodic box was simulated at 
$512^3$ resolution with a moving mesh.  At several redshift slices, 
the contents are projected onto a plane parallel to one of the box 
faces.  These projections are stacked with random horizontal and 
vertical offsets to form maps which look back through a cone of the 
sky.  The benefit to this method is that only one box must be 
simulated to create a much larger set of maps.  In our case, we have
forty maps from one simulation box.  The obvious downside to this 
technique is that a single structure in the simulation may be seen 
several times, at different orientations and stages of development, 
in different fields.  Thus the family of rare objects in the maps may 
not be truly representative.

We use the same simulations to obtain the 
convergence maps for CMB lensing, the comptonization parameter maps 
for thermal SZ, and $\Delta T/T$ maps for kinetic SZ.  These are 
discussed below for each effect.

\subsection{Lensed CMB} \label{subsec:cmb}

  We employed CMBFAST \cite{1996ApJ...469..437S} to generate the 
unlensed CMB power spectrum to $l=20\ 000$.  This power spectrum was 
used to create a Gaussian random field, which is the unlensed CMB 
map.  Lensing bends light from one part of the sky to another, so 
that the light observed from line-of-sight $\n$ in reality came from 
$\n+\delta\n$.  The real space displacement vector $\delta\n$ may be 
calculated via the Fourier space relation, $\delta\n(\mathbf{l}) = 2i 
\mathbf{l} \kappa(\mathbf{l})/ l^2$.  The convergence map for CMB 
lensing $\kappa(\n)$ is defined as a line-of-sight projection of the 
density perturbation $\delta \equiv \rho/\bar{\rho} - 1$:
  \begin{equation} \label{eqn:kappa}
    \kappa(\n) = \frac{3 H_0^2 \Omega_m}{2 c^2}  \int_0^{\chi_{\rm LS}} 
{r(\chi)r(\chi_{\rm LS}-\chi) \over r(\chi_{\rm LS})} 
{\delta(\bn) \over a} d\chi.
  \end{equation}
  The integral is over the comoving radial coordinate $\chi$.  Here 
$\chi_{\rm LS}$ is the coordinate of the CMB's surface of last scattering 
at recombination, the line element radial function $r(\chi)$ is simply 
$\chi$ for a flat universe, 
$\Omega_m$ is the matter density parameter, $\bn$ describes position 
$(\n,\chi)$, and $a$ is the expansion 
factor of the universe.  Software from \citeNP{1999PhRvD..59l3507Z} 
computes the lensed maps from unlensed maps and the convergence (See 
also \citeNP{2003PhRvD..67d3001H}).  Hydrodynamical simulations 
provided the convergence calculated from the gas density, $\delta_g$. 
Here we assume assume gas traces matter, 
which should be valid on scales of interest here. 
The primary 
CMB has Gaussian fluctuations, but lensing introduces 
non-Gaussianities by correlating the CMB with the nonlinear matter 
field.

\begin{figure*}
\begin{center}
\includegraphics[width=\widefig]{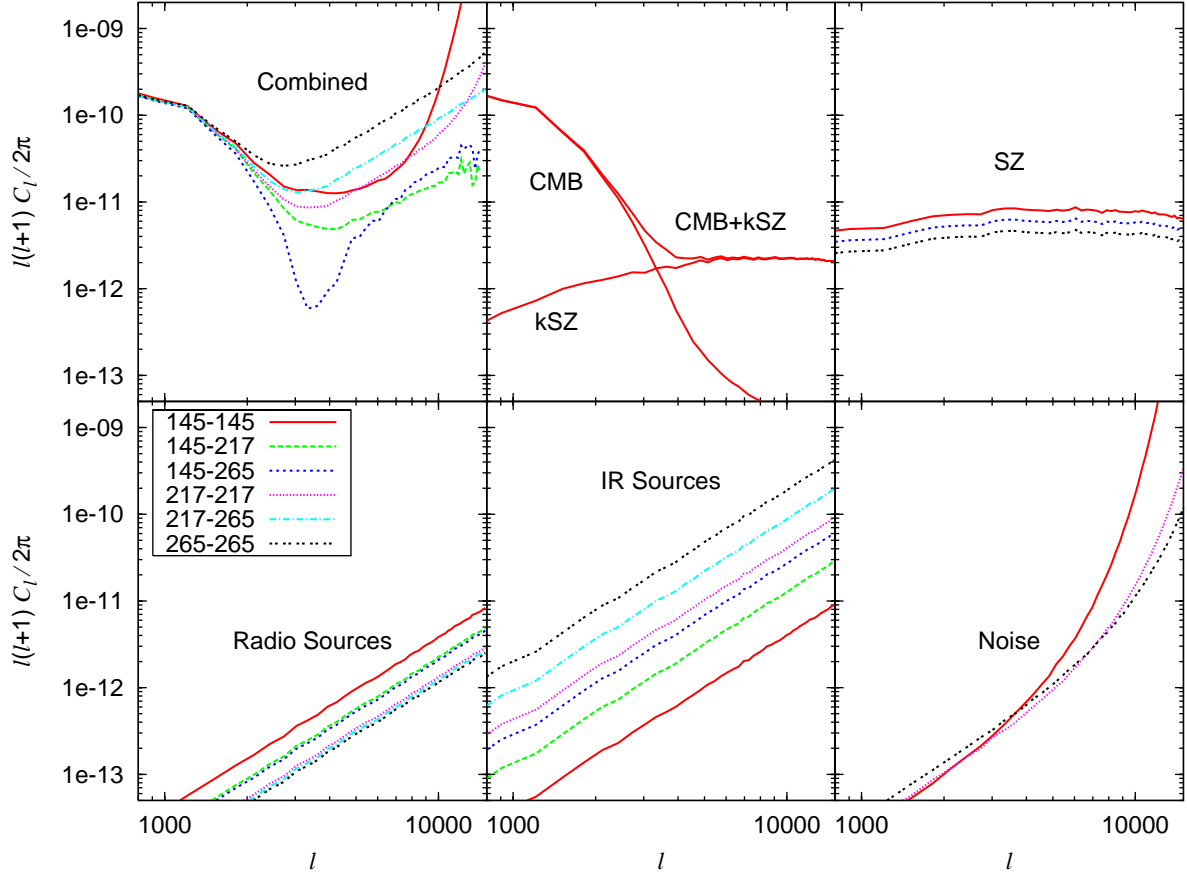}
\end{center}
\caption{Auto- and cross-power spectra from the ACT bands, in bins of 
width $\Delta l \approx 300$.  $C_l$ measures the power in 
temperature perturbation.  Line type describes the frequency 
dependence of the signals.  ``145-145'' denotes the power spectrum in 
the 145 GHz channel.  ``145-217'' denotes the cross correlation 
between the 145 GHz and 217 GHz channels.  Others correlations are 
labeled similarly.  Each signal is depicted in a separate panel.  ACT 
measures the sum of all signals and noise, shown in the panel marked 
``combined.''    At each frequency in the combined signal, the CMB 
dominates at large scales.  Small scales are dominated by 
experimental noise and infrared point sources.  Effects with a Planck 
spectrum (CMB/kSZ) do not vary with frequency.  The SZ 145 GHz-265 
GHz cross correlation is negative, and is shown in absolute value.
}
\label{fig:crosspower}
\end{figure*}

A sample lensed CMB map is shown in Figure \ref{fig:signals}. The 
lensing has negligible visible effect on the maps.  
The CMB power spectrum is shown in Figure \ref{fig:crosspower}.  
Because we examine the temperature perturbation power spectrum, any 
signal with a blackbody frequency dependence will contribute equally 
in the three bands.  This is the case for the CMB.  Since the maps 
are only $1.2^{\circ}$ on a side the  power spectrum bins 
have a width $\Delta l \approx 300$, so the acoustic peaks are not 
resolved.  The CMB is the dominant signal on scales larger than $l 
\sim 1000$.  The major features are the damping tail, where the power 
falls exponentially at $1000 \lsim l \lsim 4000$, and the contribution 
from lensing, which dominates the power at $l \gsim 4000$.  

\subsection{Sunyaev-Zeldovich effect} \label{subsec:sz}

  The SZ (or thermal SZ) effect is a change in the spectrum of the 
CMB due to scattering off of hot gas in the potential wells of large 
structures (see \citeNP{1999PhR...310...97B,2002ARA&A..40..643C} for 
comprehensive reviews).  Scattered CMB photons are preferentially 
raised in energy.  This distorts the energy 
spectrum: in the direction of the hot gas, the CMB has fewer 
low-energy photons and more high-energy photons.  The number of 
scatterings for a photon passing through a cluster is low, so the CMB 
photons do not achieve thermal equilibrium with the electrons.  
Therefore, the SZ effect does not follow a blackbody spectrum.  If 
the electrons are not relativistic then the 
frequency dependence is
  \be
  \Theta^{\rm SZ}(\n) = -2 y(\n) \left[ 2 - \frac{x/2}{\tanh(x/2)}  
\right], \label{eqn:szfreqdep}
  \ee
  where $x=h\nu/k_B T_{\rm CMB}$ and the comptonization parameter $y$ 
is given by the integral of the electron pressure along the line of 
sight:
  \begin{equation}
    y(\n) = \int \sigma_T n_e(\bn) {k_B T_e(\bn) \over m_e c^2} 
a d\chi.
  \end{equation}
  Here $\sigma_T$ is the Thomson scattering cross section, $n_e$ is 
the electron density, and $T_e$ is the electron temperature.  This 
expression is derived in the single scattering, non-relativistic 
limit.  In the low frequency 
(Rayleigh-Jeans) regime, $x \ll 1$ and $\Theta^{\rm SZ}=-2y$.   At 
roughly 217 GHz, the SZ effect is absent.  This is known as the SZ 
null.  At frequencies lower than the null, the spectrum is 
underpopulated relative to a blackbody, and the effect is seen as a 
temperature decrement.  At higher frequencies, the spectrum is 
overpopulated, and the effect is a temperature increment.  This very 
specific deviation from a thermal spectrum is SZ's signature, and is 
vital for extracting the effect. 
At high electron temperatures, the formula for the frequency dependence
needs relativistic corrections.   These corrections in particular displace
the SZ null.  However, this effect has a minor impact for the application here:
even if the electron temperature was increased everywhere to 
$T_e = 20\ 000$~K (holding the $y$-parameter constant), the SZ power spectrum
remains a minor addition to the 217 GHz band.

 A sample SZ map, showing negative Rayleigh-Jeans temperature, 
is depicted in Figure 
\ref{fig:signals}.  The SZ effect is most 
striking in clusters. The resulting signal is non-Gaussian.

The SZ effect has been studied extensively with simulations (e.g. 
\cite{2002ApJ...577..555Z,2000MNRAS.317...37D,2001ApJ...549..681S,
2001PhRvD..63f3001S,2002ApJ...579...16W,2002PhRvD..66d3002R}) 
and semi-analytic methods 
\cite{2002MNRAS.336.1256K,2000PhRvD..62j3506C}. 
There are some differences in the results between different 
simulations and halo model predictions (see \citeNP{2002MNRAS.336.1256K} 
for a discussion), but these are most 
prominent on very small scales, while in the relevant range 
around $l \sim 3000$ the agreement in the power spectrum is quite good. 
The SZ power spectrum from simulations used here \cite{2002ApJ...577..555Z}
is shown in Figure \ref{fig:crosspower}.  The 
effect of observation frequency is apparent.  The SZ effect is most 
prominent, and negative, in the 145 GHz channel, absent at 217 GHz, 
and is substantial at 265 GHz.  At very large scales, the discrete 
nature of clusters is apparent, and the power spectrum is like shot 
noise, with $C^{\rm SZ}_l$ constant.  At smaller scales, the power 
goes as $C^{\rm SZ}_l \propto l^{-2}$ for scales of $2000 \lsim l 
\lsim 10000$.  At yet higher $l$, the power falls faster.

\subsection{Kinetic Sunyaev-Zeldovich effect} \label{subsec:ksz}
The kinetic SZ effect is due to the scattering of photons off gas 
with a radial peculiar velocity.  It may be expressed as an integral 
over the density weighted gas velocity projected in the line of sight 
direction.
\be
\Theta^{\rm kSZ}(\n) = \int \sigma_T X_e(\bn) n_e(\bn) 
\frac{\mathbf{v}(\bn) \cdot \n}{c} \exp[\tau(\chi)] a d\chi.
\ee 
Here $X_e$ is the ionization fraction of the gas and $\tau(\chi)$ is 
the optical depth to 
comoving radius $\chi$.  As a Doppler shift, this fluctuation affects 
only the radiation temperature, so the spectrum of this effect is 
still blackbody. Photons scattering from electrons with radial 
velocity away from us 
show a temperature decrement and vice versa.  In our simulations, we assume
a spatially homogeneous ionization fraction $X_e$.  We discuss this 
assumption in section \ref{subsec:simcaveats}.
Figure 
\ref{fig:signals} shows an example kSZ map.  Note the association 
between the kSZ effect and SZ effect.  The same gas causes both the 
kSZ and the SZ effect, but the cross-spectrum 
vanishes because, on average, half of clusters have a peculiar 
velocity towards us, and half away from us.  The equivalent of kSZ at 
higher redshifts is 
called the Ostriker-Vishniac effect. The second order term  dominates over
the linear effect, which is strongly suppressed. This effect is thus also 
expected to be non-Gaussian.
Figure \ref{fig:crosspower} shows the kSZ power spectrum.  Like the 
CMB, kSZ is equally represented in all three channels.  
At higher $l$, we have roughly $C^{\rm kSZ}_l \propto l^{-2}$ 
for $4000 \lsim l \lsim 15000$ \cite{2004MNRAS.347.1224Z} before 
falling off at high $l$ (see \cite{2001MNRAS.326..155D,2001ApJ...549..681S}
 for other simulations
of the kSZ).
Modeling the kSZ remains quite uncertain.  It is sensitive to the 
details of reionization which are not yet well understood and the power 
spectrum is probably uncertain to an order of magnitude 
\cite{2003astro.ph..5471S}.

\subsection{Point sources} \label{sec:simps}

Two populations of extragalactic point sources are the main source of 
astrophysical contamination for the survey.  These are radio 
galaxies, brightest in the lowest frequency channel, and dusty 
galaxies, brightest in the highest frequency channel.  Dusty infrared 
point sources are the more serious contaminant.  Point sources 
dominate the secondary anisotropies in the two higher frequency 
bands, and obey Poisson statistics (assuming no correlations).  Thus 
the statistics are non-Gaussian.  In the following, we discuss each 
family of sources, point source removal by masking, and the frequency 
coherence of the source population as a whole.

\subsubsection{Radio point sources}

Synchrotron emitting point sources feature prominently in the WMAP 
maps.  They are less prominent here because of the frequency 
dependence, but still make a contribution.  We model radio sources 
using a fit to the source count model of \citeNP{2002astro.ph..4038D} 
based on \citeNP{1998MNRAS.297..117T}.
  For each flux level, we compute the number of sources per field 
according to a Poisson distribution, where the average density is 
given by the source counts.  We place the point sources without 
spatial correlation, as this is simplest, and any correlations, 
particularly for faint sources, are diminished by projection.  The 
flux of these radio sources decreases with increasing frequency, 
typically scaling as $\nu^{-0.5}$ to $\nu^{-1}$.  For our 
simulations, we use a uniform random deviate for the exponent of the 
frequency dependence for each source, and compute the flux in each of 
the ACT bands, converting to temperature units.  

The power in uncorrelated point sources is given by 
\be  \label{eqn:pspow}
C^{\rm ps}_l = \frac{1}{T_{\rm CMB}^2}  
\left[\frac{dB}{dT}\right]_{T_{\rm CMB}}^{-2} \int_0^{S_{\rm cut}} 
\frac{dN}{dS} S^2 \ dS
\ee
where the derivative of the Planck spectrum $dB/dT$ is used to 
convert into temperature units, and $dN/dS$ is the differential 
source counts.  After masking of bright sources, described below in 
section \ref{subsubsec:mask}, the radio point source power spectrum 
is plotted in Figure \ref{fig:crosspower}.   The radio power decreases 
with frequency.  Since the sources are assumed to be
spatially uncorrelated, the power spectrum is white: $C^{\rm rad}_l$ 
is constant.  Radio point sources do not dominate in any of the bands 
at any scale, but proper treatment of them is important to get 
unbiased power spectrum results.  

\subsubsection{Infrared point sources}

Dusty galaxies reprocess starlight to emit strongly in the infrared.  
To model these sources, we use source counts at 353 GHz.  
These source counts are based on a model of the optical emission of 
galaxies combined with a model for interstellar dust re-emission in 
the far infrared of absorbed optical light 
\cite{2002ApJ...570..470T}.  This model's counts are within a factor 
of $\sim 2$ of the SCUBA counts at the same frequency.  We place them 
in the fields in the same manner as the radio sources, with no 
spatial correlations.   The flux of these sources depends on 
frequency roughly as $\nu^{3.5}$, with some scatter.   Thus infrared 
sources become more prominent at higher frequency.  For our 
simulation, we chose frequency dependences of $\nu^3$--$\nu^4$.

After masking out the bright sources (section \ref{subsubsec:mask}) the 
infrared point source power spectrum is plotted in Figure 
\ref{fig:crosspower}.   The IR power increase with frequency.  Again
the power spectrum is white,
since the sources are assumed to be spatially uncorrelated.  
Infrared point sources are more contaminating than the radio sources 
in the higher frequency bands, and are the major contaminant in these 
bands until the instrument noise dominates at high $l$.  At 145 GHz, 
infrared sources are dimmer than the radio sources, but there are 
many more of them, so the power in each is about the same.  The power 
in infrared sources increases with frequency.


\subsubsection{Source Masking} \label{subsubsec:mask}

To some extent, the brightest point sources can be removed 
individually by thresholding, frequency modeling, and masking.  We 
explore this procedure.  At what level may we reasonably (and 
conservatively) detect and remove point sources?  What is the impact 
on a power spectrum analysis?  Is information from surveys outside 
the ACT survey helpful in constraining the point source contribution? 

For simplicity in our simulations, we do not attempt to extract or 
mask point sources individually.  Rather we impose a flux limit in 
our simulation, imagining that the most offending point sources have 
been already excluded. 

Our strategy for identifying point sources focuses on the 145 and 265 
ACT bands, where radio and infrared sources are most prominent, 
respectively.  
We spatially filter the maps to emphasize point sources and 
de-emphasize the CMB and detector noise.  As a side effect, this also 
enhances SZ clusters in the maps.  The filter we use is the 
reciprocal of the channel's power spectrum (Figure 
\ref{fig:crosspower}), but any similar broad filter would work.  We 
set a flux thresholds, and identify sources.

A $+5\sigma$ threshold at 145 GHz identified radio point sources down 
to 4 mJy.  At 145 GHz SZ clusters are negative, so there is no chance 
to confuse SZ sources as point sources.  (We ignore the possible 
correlation between sources and clusters, which may lead to the 
unintended elimination of some of the 
SZ clusters and thus reducing the SZ signal.)  
At 265 GHz, a $+5\sigma$ threshold identifies infrared point sources 
to 5.5 mJy.  This threshold is sufficiently high that only the very 
brightest SZ clusters could be misidentified.  Source shape and 
frequency information should eliminate any confusion.  Thus we adopt 
these flux cuts in our simulations.

As point sources are masked out, we reduce the survey area.  The 
masked area can be calculated from the source counts.  We assume 
the largest beam ($1.7$ arcmin) as the rough scale of a mask.  The 
fraction of the survey remaining unmasked after $N$ sources are 
excised is roughly $(1-({\rm mask})/4\pi f_{\rm sky})^N$, assuming 
masks can overlap.  The radio and point source cuts described above 
will remove about 2 percent of the survey area.

Might we make a more aggressive flux cut to reduce the power in power 
sources, using multifrequency information or a different point source 
detection scheme?  Yes, but to do so is not always useful.  In a 
power spectrum analysis, a measured signal has variance like
\be
{\rm var}(C^{\rm signal}_l)   \propto \frac{(C^{\rm 
signal}_l+C_l^{\rm ps}+C_l^{\rm noise})^2}{f_{\rm sky}}. 
\label{eqn:psvar}
\ee
The purpose of masking sources is to reduce $C_l^{\rm ps}$, thus 
reducing the variance of our measured signal.  Note, however, that 
masking also reduces $f_{\rm sky}$, which can serve to increase the 
variance if we mask too many sources.  We shall see that for infrared 
sources, even simple methods for identifying point sources identifies 
so many sources that masking them all would be counterproductive.

This can be understood visually in Figure \ref{fig:channels}.  A 
simple multifrequency combination can readily identify infrared point 
sources below the 5 mJy flux cut at 265 GHz applied that image.  For 
example, compare the 265 and 145 GHz channels and imagine masking 
every visible point source.  The bulk of the survey would be consumed 
by the mask.

As a more concrete example, lowering the flux cut on infrared sources 
from 5.5 mJy to 1.6 mJy at 265 GHz removes $\sim 35$ percent of the 
survey area.  Such a flux cut reduces the power in such sources by 
$\sim 20$ percent.  For a signal whose variance is dominated by faint point 
sources, this aggressive flux cut improves the variance only 
slightly.  For a signal comparable in amplitude to the point source 
noise, the variance actually worsens.  At these flux levels the power 
in point sources is not dominated by a few bright sources, but by a 
mass of background sources.  

Thus it may be sufficient, or even advantageous, to use a more 
conservative point source removal.  Whether or not to aggressively 
pursue point source removal depends on the particular application.  
Given a model for the signal and the source counts, one could in 
principle minimize the variance using expressions \ref{eqn:pspow} and 
\ref{eqn:psvar}.  Since we are examining signals over a wide range in 
$l$, and the relative power in point sources and other signals varies 
greatly, we will simply use the more conservative cut for infrared 
sources.  

The differential counts for radio sources at the level of our flux 
cut are shallower, so masking is useful.   It would also be helpful 
to compare the ACT survey at 145 GHz to radio surveys covering the 
same area, to assist in identification and masking of radio sources.  
This could reduce the variance caused by radio sources, if such 
information were available.  However, we will proceed as if the 
fields from ACT are the only measurements of those sources available 
to us.
Information from outside surveys to identify and mask offending radio 
point sources is helpful, but not necessary for a power spectrum 
analysis with ACT. 

\subsubsection{Frequency Coherence}
For some purposes, it is useful to treat all sources as a single population.
In this case, the details of the multiple source populations, frequency 
dependence, and scatter, may all be subsumed under a set of correlation 
coefficients which describe the frequency coherence across bands.
If the correlation coefficient for point sources between two bands is unity, 
then one band is a scaled version of the other.  In our point source model, 
the 
correlation coefficient between the 217 and 265 GHz bands is $0.99$, close
to unity.  
The deviation from unity is dominated by scatter in the scaling of the 
IR sources.  However, the point source population in the 145 band has a 
significant contribution from radio sources.  We find the correlation 
coefficient between 145 and 217 GHz is $0.83$.  Between 145 and 265 GHz the
 correlation  is $0.76$.  Ignoring this frequency incoherence has serious 
consequences when 
estimating the power spectrum, as we note below.

\subsection{Detector noise}
We add Gaussian distributed detector noise to these simulated fields. 
 In reality, the cosmic signals will be convolved with the beam of 
the telescope, and the detector will add white noise.  Here we find 
it more convenient to take the mathematically equivalent approach of 
de-convolving the beams from the three channels.  Thus the cosmic 
signals are untouched and the noise is correlated, with noise power 
increasing at small scales.  Assuming the noise and Gaussian beams in 
Table \ref{tab:actspecs}, the power spectrum of noise is given by:
\be
C_{l \alpha}^{\rm noise} = (\sigma_\alpha  b_\alpha)^2 \exp\left( 
\frac{(l{ b}_\alpha)^2}{\sqrt{8 \log 2}}  \right)
\ee
Where $\alpha$ stands for the channel, $\sigma_\alpha$ is the RMS 
noise per beam-sized pixel, and the beam full width at half maximum $ 
b_\alpha$ must be converted to radians.  Figure \ref{fig:crosspower} 
shows the noise power spectrum in the three channels.  At large 
scales the noise is similar in the three channels.  At smaller 
scales, the noise dominates in the 145 GHz channel first, then at 217 
GHz, then 265 GHz.

\subsection{Caveats and other signals not included} 
\label{subsec:simcaveats}

Here we list several caveats and further possible improvements 
that we did not include in our simulations.  One caveat is that the
simulations have $\sigma_8 = 1$ normalization, which is at the high 
end of what 
is currently favored.  In particular, this may make the SZ fluctuations 
larger than in reality, since SZ is rapidly increasing with the 
matter power spectrum amplitude.

The hydrodynamical simulations we incorporate into our mock survey 
assume a spatially uniform transition at reionization.  In reality, 
the ionized fraction will depend on where ionizing photons are 
produced, their ability to propagate, and their efficiency to ionize. 
 Since these quantities probably vary from place to place, 
reionization is not likely to be uniform.  Modeling of this so-called 
patchy reionization is difficult, and the impact remains somewhat 
uncertain.  Thus we have not included it in this analysis.  However, 
its impact may be significant: in recent modeling by 
\citeNP{2003astro.ph..5471S}, patchy reionization typically increased 
the power in the kSZ effect by an order of magnitude (see also 
\citeNP{2000ApJ...530....1M}). 

The modeling of point sources is still quite uncertain.  Source 
number counts, clustering, frequency dependences and possible 
correlations with SZ all require 
improvements as better information becomes available.

Additionally, the infrared point sources may be correlated with CMB 
lensing, with this cross-correlation possibly detectable by Planck 
\cite{2003ApJ...590..664S} using the 545 GHz band.  We have not 
included this effect and do not consider it.  The peak scale of the 
effect at $l \sim 150$, frequency scaling of point sources, and 
$f_{\rm sky}$ of the experiments favor Planck, not ACT, to detect 
this effect. 

We have not included any galactic contaminants in these simulations.  
Because the survey covers a small fraction of the sky, we have 
assumed that a clean enough window could be found to look out of the 
galaxy with minimum contamination or that a slowly varying galactic 
component can be projected out of the analysis.

Finally, the model of the instrument we include is simplistic.  The 
actual survey will have a nontrivial geometry, non-uniform 
noise with a $1/f$ component, and non-Gaussian beams.

\begin{figure*}
\begin{center}
\includegraphics[width=\narrowfig]{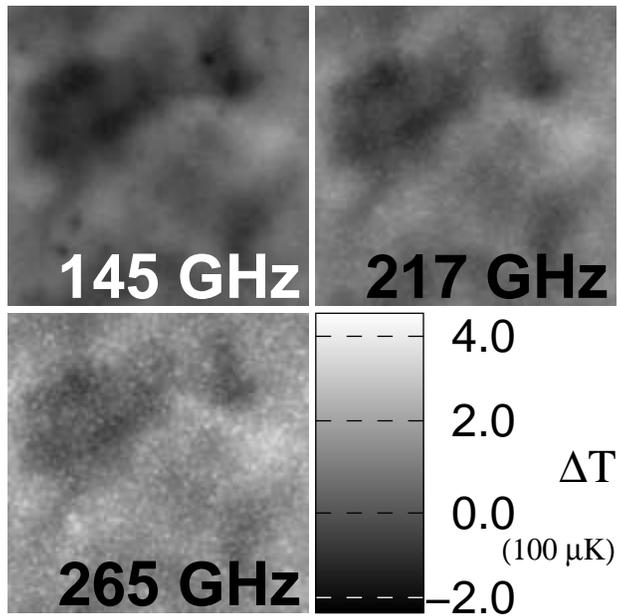} 
\end{center}
\caption{A simulated $1.2^\circ \times 1.2^\circ$ field in the bands
of the ACT telescope.  Pixel value indicates the 
difference in temperature from $T_{\rm CMB}$.  
The mean flux increases with frequency because of the 
increasing background of galaxies shining in the far infrared.  
Small scale temperature decrements at 145 GHz indicate the SZ 
effect.}\label{fig:channels}
\end{figure*}

\subsection{Summary of simulations} \label{subsec:sigsum}

We have constructed a 57 square degree mock CMB survey according to 
the specifications of ACT.  It includes five astrophysical 
components.  The CMB is a Gaussian random field, lensed by a 
convergence map from hydrodynamical simulations.  The same 
simulations produced the gas pressure (for SZ) and momentum (for 
kSZ).  Both SZ and kSZ have well defined frequency 
dependences.  SZ is negative in the 145 GHz channel, zero in the 217 
GHz channel, and positive in the 265 GHz channel.  Kinetic SZ scales 
with frequency exactly as the primary CMB.  The simulations include 
two types of point sources.  Radio sources dim with frequency and 
have their greatest contribution at 145 GHz and little contribution 
at higher frequencies.  Infrared point sources brighten with 
frequency, and are prominent in the 217 GHz and 265 GHz bands.  We 
have conservatively removed the brightest radio and infrared sources. 
 A more aggressive flux cut on radio sources can help a power 
spectrum analysis, if more information about the sources is available 
from other surveys, but we assume it is not.  A more aggressive flux 
cut on infrared sources begins to degrade the survey area, and 
depending on the application may not improve the variance in a power 
spectrum analysis.  We have to include infrared sources in any 
analysis because they are too numerous to excise.  We include 
instrument noise and beam from Table \ref{tab:actspecs}.  The 
instrument has highest resolution at 265 GHz and lowest at 145 GHz.

The assembled signals, convolved with the telescope beam and added to 
the noise, are depicted in the ACT channels in Figure 
\ref{fig:channels}.  The striking features are the primary 
fluctuations on large scales, bright SZ clusters (as a decrement) at 
145 GHz, and the substantial brightening of infrared point sources at 
higher frequencies.

The power spectra of these channels are shown in Figure 
\ref{fig:crosspower}.  The power spectrum in each of the three 
channels is dominated by the CMB on the larger scales and 
instrumental noise on the smaller scales.  At the intermediate scales 
of this survey, SZ is the most significant component in the 145 GHz 
band.  In the bands at 217 GHz and 265 GHz, infrared point sources 
are the major contribution to the point sources at intermediate 
scales.  

At this point we may draw several conclusions about these 
simulations.  The CMB is sub-dominant on a wide range of scales 
accessible to this survey.  The dominant cosmological features are 
secondary anisotropies, but these suffer from serious pollution from 
extragalactic point sources, particularly infrared sources in the 
high frequency bands.  These astrophysical components exceed the 
noise over a wide range of interest.  
In section \ref{sec:2point} we proceed with a power spectrum analysis 
and return to non-Gaussian features in \ref{sec:ng-pdf}.

\section{Power spectrum analysis} \label{sec:2point}

From our simulated maps, we seek to measure the power spectra of our 
various 
cosmic signals.  We use a minimum-variance quadratic estimator, which 
can also
be viewed as an iterative procedure to find maximum-likelihood 
solution 
\cite{1998PhRvD..57.2117B}.  
The power spectrum estimates are unbiased.  
For components that cannot be separated using 
the frequency information (formally in this case the Fisher 
matrix becomes singular and the errors infinite) it is better 
to treat them as a single component and then try to separate them 
using 
additional considerations such as  
a prior knowledge of the power spectrum shape for the signals.  
For some applications, power-spectrum 
estimators that involve the intermediate Wiener filtered
maps are minimum variance \cite{1998ApJ...503..492S}.
However for this application, they are not necessarily minimum 
variance.  Our approach here is a generalization of those 
treatments and should improve upon those estimators.
 As an added bonus, the effects of imperfect cleaning of one 
component 
on the mean and errors of other components are automatically 
included, for the estimator is unbiased and minimum variance 
(considering the imperfect cleaning, see 
\citeNP{2000ApJ...530..133T}).
 For non-Gaussian signals, the estimator is unbiased, but not minimum 
variance.  In this section, we discuss this estimator and its 
relation to the Wiener filter.  We discuss the impact of scatter in 
the frequency dependence of signals.  We measure the power spectra in 
several cases, making different assumptions about our knowledge of 
the point source frequency dependence.  Finally, we discuss the 
importance of cleaning out kSZ in determining the amplitude and slope 
of primary fluctuations.

\subsection{Multifrequency estimator} \label{subsec:estimator}
\def\t{\tilde}

In this section we describe the power spectrum estimator we employ. 
We will reference a particular Fourier mode in the flat sky 
approximation by its wavevector $\mathbf l$.  We use $l$ to denote 
the magnitude of this wavevector.  We organize our data into a vector 
$\mathbf{e}=\{ e^\mathbf{l}_\alpha \}$.  The subscript $\alpha$ 
refers to which band the data comes from.  We will reserve Greek 
indices to refer to the channels of the detector, 
$\alpha\beta\gamma$\dots.  We say that the data is given by the sum 
of the response of the instrument to the signals and the noise.  
Dropping the mode index, we write the data in channel $\alpha$ as 
$e_\alpha$, the signal as $s_\alpha$, and the noise as $n_\alpha$.   
For a particular multipole, we characterize the data, signals, and 
noise by their covariances $\langle e_\alpha e_\beta^* \rangle = 
C_{\alpha\beta} $, $\langle s_\alpha s_\beta^* \rangle = S_{\alpha 
\beta}$ and $\langle n_\alpha n_\beta^* \rangle = N_{\alpha\beta}$.  
Note that we could additionally assume the noise between channels is 
uncorrelated, but it is not necessary to do so.  We assume that the 
signal and noise are uncorrelated, and that the signal depends on a 
set of parameters $\bi{Z} = Z_p$.  Thus, restoring the multipole 
dependence, the covariance matrix is
\be
\bi{C} = \langle \bi{e e}^\dag \rangle = C^l_{\alpha \beta} 
\delta_{\bi{ll}'}
\ee
where 
\be
C^l_{\alpha\beta} = S^l_{\alpha \beta}(\bi{Z}) + N^l_{\alpha \beta}.
\ee
The covariance matrix $\mathbf{C}$ has both frequency channel and and 
multipole indices, and we have assumed that different multipoles are 
independent, hence the $\delta_{\bi{ll}'}$.

In a Gaussian model, the likelihood function for the data is
\begin{equation}
L(\bi{e} )=(2\pi)^{-N/2} \det(\bi{C})^{-1/2} \exp(-{1 \over
 2}
\bi{e}^{\dag} \bi{C}^{-1}\bi{e}).
\label{lik}
\end{equation}
A second order expansion in $Z_p$ gives
\bea \nonumber
\ln L(\bi{Z}+\delta \bi{Z})&=& \ln L(\bi{Z})+\sum_{p} 
{\partial \ln L(\bi{Z}) \over \partial Z_{p}}\delta Z_{p} \\ 
&& + {1 \over 2} \sum_{pp'} {\partial^2 \ln L(\bi{Z}) \over
\partial Z_{p}\partial Z_{p'}}\delta Z_{p} \delta Z_{p'},
\eea
where
\begin{eqnarray}
-2{\partial \ln L(\bi{Z}) \over \partial Z_{p}} &=&
\trace (\bi{e}^{\dag} \bi{C}^{-1} {\partial \bi{C} \over \partial 
Z_p} \bi{C}^{-1}\bi{e} -\bi{C}^{-1} {\partial \bi{C} \over \partial 
Z_p}  ) \nonumber \\
-{\partial^2 \ln L(\bi{Z}) \over
\partial Z_{p}\partial Z_{p'}}&=&\bi{e}^{\dag} \bi{C}^{-1}
{\partial \bi{C} \over \partial Z_p}
\bi{C}^{-1}{\partial \bi{C} \over \partial 
Z_{p'}}\bi{C}^{-1}\bi{e}\nonumber \\
&&-{1 \over 2} \trace (\bi{C}^{-1}{\partial \bi{C} \over \partial 
Z_p}\bi{C}^{-1} {\partial \bi{C} \over \partial Z_{p'}}).
\end{eqnarray}
The ensemble average of the second expression is the Fisher matrix
\begin{equation}
F_{pp'}=
{1 \over 2} \trace (\bi{C}^{-1} {\partial \bi{C} \over \partial 
Z_p}\bi{C}^{-1}{\partial \bi{C} \over \partial Z_{p'}}).
\label{fisher}
\end{equation}
Note that for a given value of $l$, the trace accounts for the sum 
over $(2l+1)f_{\rm sky}$ modes.
At the maximum likelihood value the first derivative of the likelihood
function vanishes, so we use the Newton-Raphson method to find the 
zero
of the derivative. This leads to the minimum variance quadratic
estimator for the parameters $\bi{Z}$
\cite{1997MNRAS.289..285H,1997PhRvD..55.5895T,1998ApJ...503..492S}.
\begin{eqnarray}
\label{wffll}
\hat{Z}_{p}&=&{1 \over 2}\sum_{p'}F^{-1}_{pp'}[\bi{e}^{\dag} 
\bi{C}^{-1}   {\partial \bi{C} \over \partial Z_{p'}}   
\bi{C}^{-1}\bi{e}-b_{p'}]
\nonumber \\
b_{p}&=& \trace [\bi{C}^{-1} {\partial \bi{C} \over \partial Z_{p}}   
 \bi{C}^{-1} \bi{N}].
\end{eqnarray}
We use $\hat Z_p$ to denote the estimate of $Z_p$.

In this formalism, the covariance matrix of the estimated parameters, 
written $\mathsf C_{pp'}$, is given by the inverse Fisher matrix:
\be
{\mathsf C}_{pp'} = \langle \hat Z_p \hat Z_{p'} \rangle = F^{-1}_{pp'}
\ee

This technique requires knowledge of the covariance $\bi{C}$.  We may 
estimate this by taking a smoothed version of the data.  If the 
covariance is incorrect, the estimator is still unbiased, but the 
error bars will be incorrect.

The parameters we desire to estimate are the power spectra of the 
signals, defined in a way which does not depend on the frequency 
channels of our instrument.  
We denote our signal with $\mathbf{s} = s^\mathbf{l}_m$.  Here $m$ 
indicates which signal we are referring to.  We will also use the 
(cross) power spectrum of signals, denoted $S^l_{mn} = \langle  
s^\mathbf{l}_m s^\mathbf{l'}_n \rangle \delta_{\bi{ll}'}$.  We 
introduce the frequency response $\mathcal{R}_{\alpha \beta mn}$ to 
relate the spectra of the signals in the channels to the spectra of 
the signals $S^{l}_{mn}$, however we have chosen to define them.  
Note that $S^{l}_{mn}$ is symmetric in $m$ and $n$, so only one 
ordering of each pair is required. 
Then the cross power spectrum from the combination of channel 
$\alpha$ and channel $\beta$ will be given as:
\be
S^l_{\alpha \beta} = \sum_{mn} \mathcal{R}_{\alpha \beta mn} S^{l}_{mn}.
\ee
We discuss the frequency response in more detail below.  To estimate 
the power spectrum of the signals we set $Z_p = S^{l}_{mn}$. Note 
that in this case the index $p$ enumerates each combination of $lmn$. 
 In this case the appropriate 
derivative of the covariance is given by
\bea
\label{eqn:woshape}
&&   {\partial C^{l'}_{\alpha \beta} \over \partial Z_p} =
   {\partial C^{l'}_{\alpha \beta} \over \partial S^{l}_{mn}} =  
\sum_{m'n'} \mathcal{R}_{\alpha \beta m' n'} \delta_{ll'} \\ \nonumber
&& \  \times \left[\delta_{mm'}\delta_{nn'} + 
\delta_{mn'}\delta_{nm'} - 
\delta_{mm'}\delta_{nn'}\delta_{mn'}\delta_{nm'} \right] 
\eea
The final expression in the brackets is shorthand for:
\be
[\dots] = \left\{
\begin{array}{l l l}
  1 & \parbox{0.5\columnwidth}{if ($m=m'$ and $n=n'$) or ($m=n'$ and 
$n=m'$),}\\
  0 & \mbox{otherwise}.
\end{array} \right.
\ee

If we know the shape of the cross spectra and wish to estimate the 
amplitude, we can write the signal power in the bands as 
\be
S^l_{\alpha \beta} = \sum_{mn} \mathcal{R}_{\alpha \beta mn} A_{mn} 
S^{l}_{mn}
\ee
where $A_{mn}$ encodes the information about the amplitude, and 
$S^{l}_{mn}$ is the shape.  If we want to estimate this amplitude, 
then $Z_p = A_{mn}$.  Thus the derivative of the covariance is given 
by
\bea
\label{eqn:wshape}
 &&{\partial C^l_{\alpha \beta} \over \partial Z_p} = 
   {\partial C^l_{\alpha \beta} \over \partial A_{mn}} =
   \sum_{m'n'} \mathcal{R}_{\alpha \beta m' n'} S^{l}_{mn} \\ \nonumber
&& \ \times \left[\delta_{mm'}\delta_{nn'} + \delta_{mn'}\delta_{nm'} 
- \delta_{mm'}\delta_{nn'}\delta_{mn'}\delta_{nm'} \right] 
\eea

If we know the shape of some signals, but not others, we may use the 
two expressions (\ref{eqn:woshape}) and (\ref{eqn:wshape}) in 
combination. In these expressions we have assumed that the power 
spectra of 
the components are unknown, but that their frequency dependence is 
known (and hidden in 
$\mathcal{R}_{\alpha \beta mn}$ coefficients). If the parameters which 
determine the frequency scaling are not known then 
one can generalize this further, add these parameters to the 
estimate, and maximize the likelihood over them as well, provided the 
Fisher matrix does not become singular.  The procedure of taking 
derivatives of the covariance matrix is similar, so we do not give 
such expressions here. 

\subsection{Relation to Wiener filter}

If all signals have a spatially uniform frequency dependence, then 
$s^\bi{l}_{\alpha} = \sum_m R_{\alpha m} s^\bi{l}_m$ for all 
$\bi{l}$.  The correlation coefficient of modes between frequency 
channels will be unity (neglecting noise).
In this case we can factor the frequency response into a simpler form:
\be
\mathcal{R}_{\alpha \beta mn} =  R_{\alpha m} R_{\beta n} 
\label{eqn:splitR}
\ee
We may put the components into a vector $\bi{R}_{m}=R_{\alpha m}$.  
This allows us to compactly express the Wiener-filtered estimates of 
the signal maps, which minimize the variance in reconstruction 
errors.  The same expression holds for all modes, so we drop the mode 
index $\bi{l}$.  The Wiener estimate is
\bea
\hat{s}_m &=& \sum_{m'} S_{mm'}y_{m'}  \\
y_m &=& \bi{R}_{m}^\mathsf{T}\bi{C}^{-1}\bi{e}.
\eea

The $\hat{s}_m$ are the Wiener filtered estimates of all signal maps. 
 From these estimated maps of the signals, we can compute the power 
spectrum, taking into account the contamination from the other 
signals to the Wiener filtered maps.  In the case of spatially 
uniform frequency dependence, this procedure is identical to the 
multifrequency estimate in the previous section.  The Wiener estimate 
requires the power spectrum $S_{im}$, or we could use a power 
spectrum estimate and iterate.  

For signals with non-uniform frequency dependence, it may not be 
possible to write a vector $\bi{R}_{m}$ which satisfies 
(\ref{eqn:splitR}).  A spatially-averaged frequency dependence may 
still be used to produce a Wiener-type estimate, but it is not 
guaranteed to be an unbiased or a minimum variance reproduction of 
the signal.  In addition, the power spectrum obtained from these maps 
can be substantially biased.  This is because a power spectrum 
estimate which uses a Wiener filtered map as an intermediate step 
does not account for the scatter in the frequency dependence.  If 
this scatter is substantial, the Wiener filter may be inappropriate.  
(See \citeNP{2000ApJ...540....1C} for making minimum variance maps of 
signals with good frequency coherence in the presence of incoherent 
foregrounds.  To make a map of an incoherent signal, like a point 
source template, this approach may not be minimum variance.  See 
\citeNP{2000ApJ...530..133T} for a generic model of frequency 
incoherence.)

\subsection{Frequency dependence and scatter} \label{subsec:templates}

To apply the multifrequency filter, one needs the noise power spectra 
and signal frequency dependences.  We assume the noise power spectrum 
in the three channels is well known. 
The frequency scalings of CMB and kSZ signals are constant in 
temperature units, so the relevant components of $\mathcal{R}_{\alpha 
\beta mn}$ are unity.  For SZ, the frequency dependence is from 
equation \ref{eqn:szfreqdep}.  The frequency dependence of point 
sources are poorly known.  Significant modeling or outside 
observations are needed to fill in the remaining components of 
$\mathcal{R}_{\alpha \beta mn}$.  These cannot be {\em completely} determined 
from ACT alone in an unbiased way. If they are added to the parameter 
vector $Z_p$ without any additional assumptions, the Fisher matrix 
becomes singular.  This means these parameters are completely 
degenerate with the power spectra: there exists a family of power spectra and 
frequency dependences which reproduces the covariance, and it is 
impossible to distinguish between individual members of this family.  

It is possible, however, that for given models of the point source 
emission, the parameters of the model may be added to the estimate 
without forcing the Fisher matrix to be singular.  Additionally, we 
can use the high $l$ part of the spectrum, which is dominated by 
point source emission, to help constraint frequency dependence, at 
the expense of biasing our estimator.  We use this in the
following section.

In our simulations, the frequency scaling differs substantially from 
point source to point source.  The correlation coefficient between 
modes at different frequencies is less than unity, so the cross 
spectrum is less than predicted from the auto-powers.  It may be 
senseless to produce a Wiener filtered signal estimate based on a 
spatially averaged frequency dependence.  We do not use a spatially 
averaged frequency dependence in our power spectrum estimations, for 
we find it produces greater than $2\sigma$ biases in the power 
measurement.  The conclusion is that it is crucial to address the 
scatter in the frequency dependence of point sources.

ACT and similar experiments will identify a population of bright 
sources, which may be used to study their frequency scaling.  
One should apply this information to a power spectrum analysis with some 
caution, because the sources which contaminate the power spectrum 
analysis are dimmer than the population which is convenient to study. 
 This may mean that contaminating sources are a different population, 
more distant, younger and less evolved, or some combination.  Dust 
emission is complicated, and the emission spectra of the point 
sources likely depends on the relative abundances of the dust species 
present.  If the populations are different, it will be difficult to 
build an accurate model of the emission from these sources.  In 
addition, bright sources may result from the confusion of two dimmer 
sources.  Measurements of the scatter of source frequency dependence 
is then difficult to interpret.

High-resolution interferometric measurements would help in studying 
the source population.  In particular, 
ALMA\footnote{http://www.alma.nrao.edu/index.html}  has frequency 
coverage in the ACT bands.  It should have sufficient sensitivity to 
pick out sources which are confusing ACT, and sufficient angular 
resolution to resolve them.  As mentioned, masking infrared sources 
to a low flux cut is not very promising, but ALMA would be useful as a tool to 
explore the frequency dependence and scatter of sources.

\subsection{Measured signal power spectrum}  \label{subsec:measurepow}
We measure the power spectrum in several cases assuming differing 
prior knowledge.  Our list of cases is not exhaustive, but 
illustrates techniques one might explore.  We always assume the 
frequency dependence of CMB, kSZ, and SZ.  We enumerate our 
assumptions, and the parameters to estimate in each case.

The first two cases assume we have perfect knowledge of the point 
source frequency scaling and scatter.
\begin{description}
\item {\em Case 1.}  We assume we know the frequency dependence of 
the point sources. The parameters to estimate are the power spectra, 
in bins, of CMB$+$kSZ, SZ, IR and radio point sources.
\item {\em Case 2.}  We assume we know the power spectrum shape and 
frequency dependence of the point sources.  The parameters to 
estimate are the binned power spectra of CMB$+$kSZ and SZ, and 2 
amplitudes, for the IR and radio power spectra.
\end{description}
The next two cases assume nothing about the point source frequency 
dependence.
\begin{description}
\item {\em Case 3.}  We assume we know the shape of the point 
sources, and that the CMB$+$kSZ makes no contribution at $l>15000$.  
The final condition ensures a non-singular Fisher 
matrix.  We treat the point sources as a single population.  
Parameters to estimate are the binned power spectra of CMB$+$kSZ and 
SZ, and 6 amplitudes, corresponding to the point source amplitude in 
the cross-correlations between the bands.  This is equivalent to 
estimating the frequency dependence of the sources.  
\item {\em Case 4.}  Like case 3, but without assuming the shape of 
the point source power spectrum.  We treat point sources as a single 
population, assuming that the scatter in the frequency dependence is 
constant as a function of scale, but we do not assume what it is.  
Parameters to estimate are the binned power spectra of CMB$+$kSZ and 
SZ, the 3 binned auto-power spectra of point sources in the three 
bands, and 3 correlation coefficients, which describe the scatter in 
the frequency relation.
\end{description}
The final two cases separate kSZ from the CMB using a template for 
the kSZ power spectrum.
\begin{description}
\item {\em Case 5.}  Like case 2, including point source frequency 
and power spectrum shape information, but with two more assumptions.  
First, we assume the shape of kSZ power spectrum.  Second, we assume 
the CMB makes negligible contributions to the power for $l>6000$.  
The second additional assumption prevents a singular 
Fisher matrix.  The parameters to estimate are the binned power 
spectra of CMB and SZ, and 3 amplitudes, for the kSZ, IR, and radio 
power spectra.
\item {\em Case 6.}  Like case 3, including only point source power 
spectrum shape information, but with the additional assumptions of 
case 5.  Parameters to estimate are the binned power spectra of 
CMB$+$kSZ and SZ, 6 point source amplitudes, and the amplitude of the 
kSZ spectrum.
\end{description}
If the parameters we choose to estimate can faithfully represent the 
real world, our estimate is unbiased.
However, in any of our cases, an incorrect assumption can bias the estimate.  
Note that case 4 requires the least outside information.

Where we use a frequency dependence, we use 265 GHz as our reference 
frequency for SZ and infrared sources, and 145 GHz for radio sources. 
 This choice is a convention, and has no bearing on the quality of 
the estimate.

\begin{figure*}
\begin{center}
\includegraphics[width=\widefig]{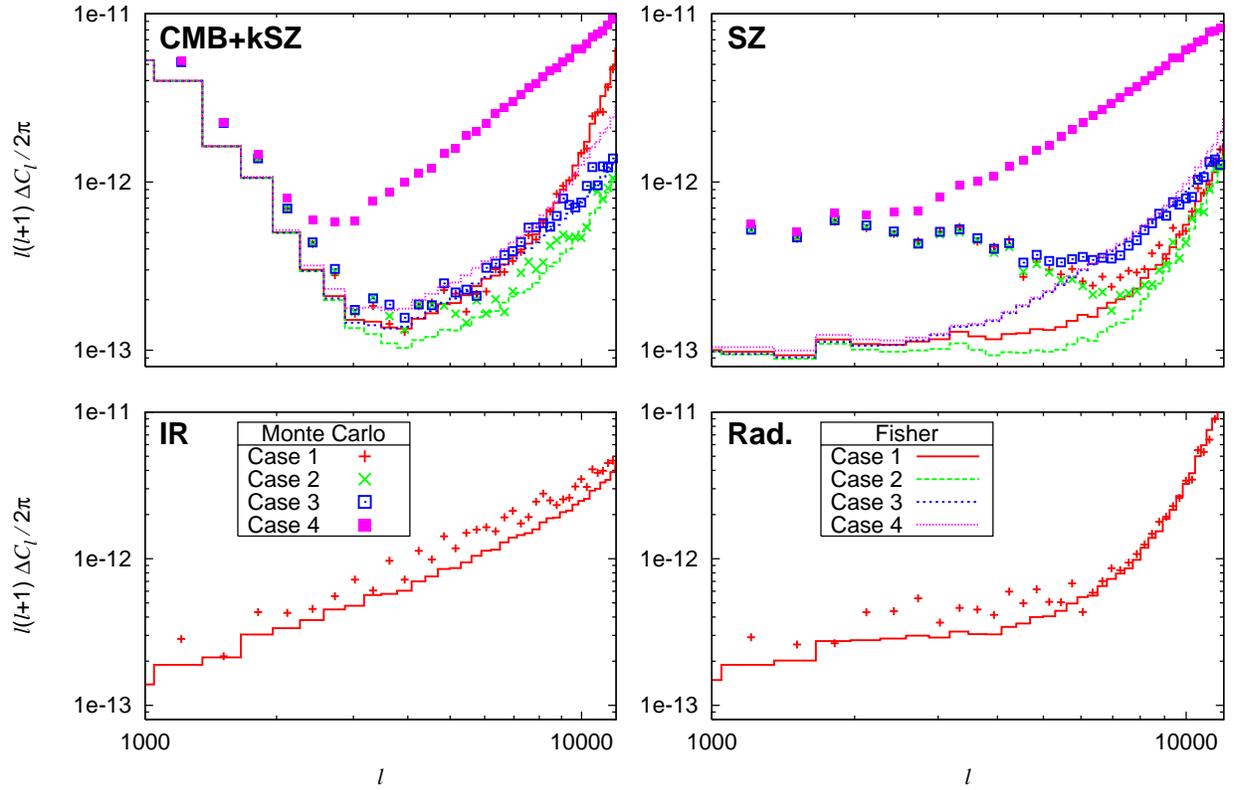}\\
\end{center}
\caption{
The errors on the estimated power spectrum.  Lines indicate errors 
derived from the Fisher matrix: $(F_{pp}^{-1})^{1/2}$.  Symbols 
indicate the measured standard deviation from 40 realizations.  
The errors in case 5 are similar to case 2, and the errors in case 6 
are similar to case 
3.  The bottom 
panels show only case 1.  For cases which assume the shape of the point 
source power spectrum, plotting the errors as a function of $l$ is 
not possible. 
\label{fig:err}
}
\end{figure*}

We estimate the covariance on the measured power spectra in two ways. 
 First we use the inverse Fisher matrix, so the error on the 
estimate of power bin $p$ is  $(F_{pp}^{-1})^{1/2}$.  Second, we computed 
the covariance from the estimates on our 40 realizations.  In Figure 
\ref{fig:err}, we show the errors on these power estimates as a 
function of $l$.  In the regime where instrumental noise dominates (at 
sufficiently high $l$), 
the inverse Fisher matrix provides a good estimate of the errors.  
Elsewhere, we find the inverse Fisher matrix underestimates the 
error, due to the non-Gaussianities.  Experiments of this 
type will likely need to rely on Monte Carlo simulations for their errors.

\begin{figure*}
\begin{center}
\includegraphics[width=\widefig]{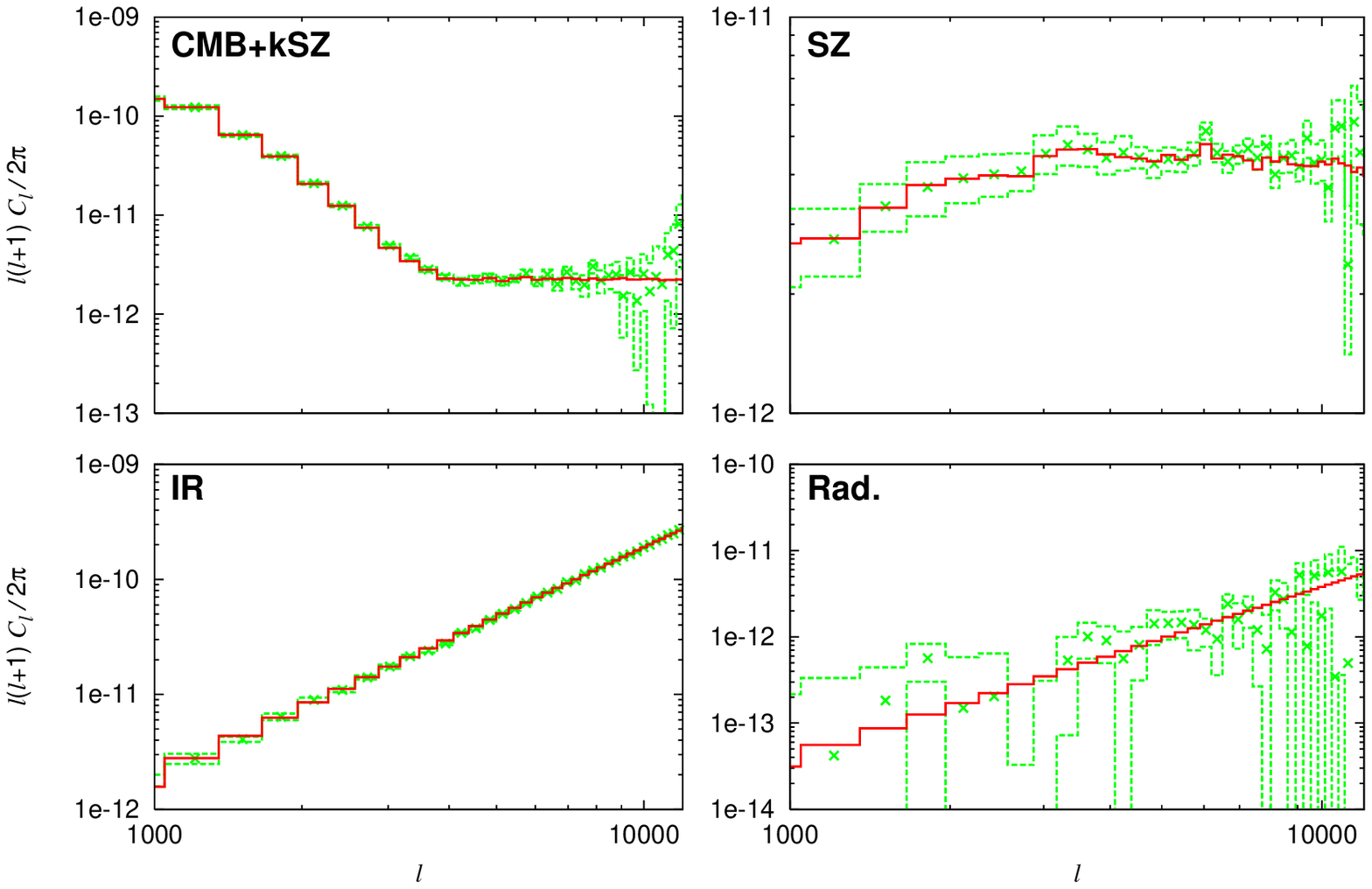}\\
\end{center}
\caption{
Estimates of the power spectra in case 1, where we assume frequency 
information, but no information about the shape of any power 
spectrum.  The solid line is the true value of the power 
spectrum.  The $\times$ symbols show the estimated value of the 
power spectrum.  The dashed lines show 1 standard deviation 
above and below the estimate, evaluated over our 40 realizations.
This estimate is unbiased.
The width of the bins is $\Delta l \approx 300$.
\label{fig:estpower}
}
\end{figure*}

We now discuss each case in turn.  The assumption in case 1 and 2 of 
point source frequency knowledge is probably dangerous if we do not 
have such knowledge.  A poor model will bias the power spectrum in a 
way which may be hard to predict.  On the other hand, if we do have a 
good understanding of the source population (from ALMA perhaps), 
cases 1 and 2 represent ideal situations.  For case 1, the measured 
power spectrum is shown in Figure \ref{fig:estpower}.  In this plot 
we use the errors evaluated over our independent realizations.  
Except for radio sources, we recover the power spectra well.  This 
provides information about the shape of the power spectrum, which we 
have not assumed.  Our result in case 1 is unbiased.

We compared the inverse Fisher matrix and the covariance matrix 
from the realizations.  In case 1, the inverse Fisher matrix is
diagonal in $l$.  However, we find that the actual covariance matrix 
is not.  The covariance between the CMB+kSZ power bins is nearly diagonal in 
this case, but to $l$ of a few thousand, the SZ power bins are strongly 
correlated.


For case 2, assuming the shape of the point source power spectrum 
improved the errors on the CMB+kSZ power spectrum by a factor of 
$\sim 4$ relative to case 1.  The errors on SZ are marginally 
improved.  The power spectrum bins are somewhat correlated.  In case 
2, the error in the power spectrum amplitude in IR and radio point 
sources is $\Delta C_l/C_l = 5.0 \times 10^{-3}$ and $\Delta C_l/C_l 
= 0.20$ respectively.  Recall that this case combines all the 
information on the point source spectra into two amplitudes.  The 
inverse Fisher matrix underestimates this error by a factor of 2.  We 
do not plot case 2 because it is unbiased and similar to case 1, with 
smaller error bars, although the bins are more correlated.

\begin{figure*}
\begin{center}
\includegraphics[width=\widefig]{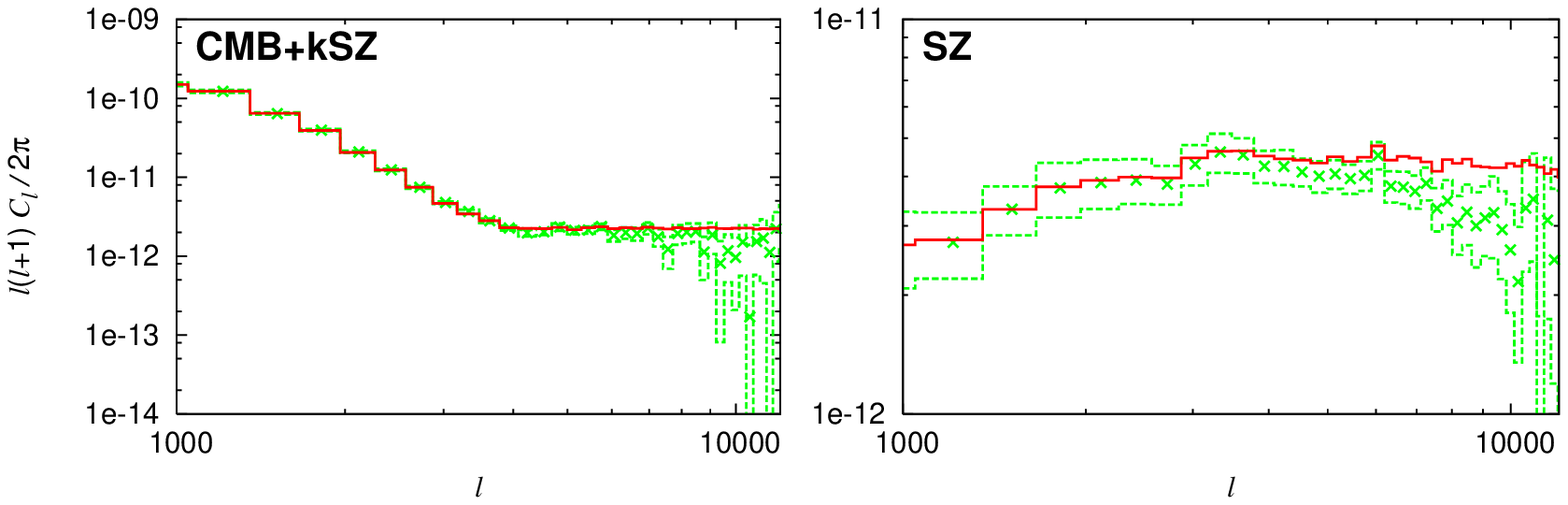}\\
\end{center}
\caption{
Estimates of the power spectra in case 3, where we assume the shape 
of the point source power spectrum, but assume nothing about frequency
dependence. 
\label{fig:estpower-case3}
}
\end{figure*}

Case 3 does not depend on prior frequency information, but has its 
own problems.  Case 3 depends on an incorrect assumption that 
only point sources contribute at high $l$, while kSZ 
continues to have some small contribution at these $l$.  Thus the power 
spectrum we deduce for point sources is slightly high, and this 
biases our measurement somewhat. This is relatively independent of 
the $l$ cutoff.  We plot the measured power spectrum in Figure 
\ref{fig:estpower-case3}, where the bias is apparent.  The power 
spectrum bins are also somewhat correlated.  However, we needed no 
model for the point source frequency dependence.  Our model is biased 
by as much as 1$\sigma$ at high $l$, but the bias is probably under 
better control than it would be for a model of frequency dependence, 
because at least we know in which direction our estimate is biased.

\begin{figure*}
\begin{center}
\includegraphics[width=\widefig]{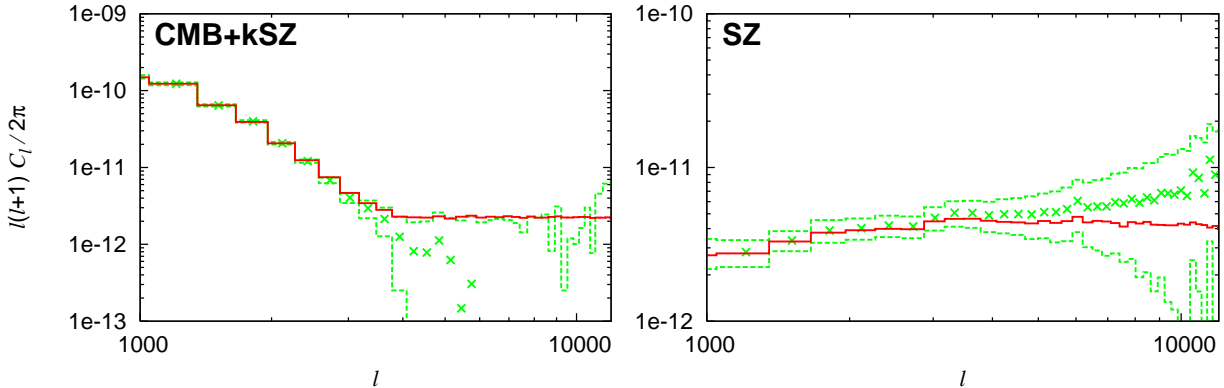}
\end{center}
\caption{Estimates for case 4.  Assuming no information about the 
power spectrum shape or 
frequency dependence of the point sources, we lose the ability to 
reliably detect kSZ.  The estimate for SZ is biased, but the 
statistical errors are large.  The bins are highly 
correlated.}\label{fig:wo-shape-no-freq}
\end{figure*}

Case 4 assumes less about the point sources than case 3, because it 
does not assume a shape for the power spectrum.   We plot the power 
spectrum of CMB+kSZ and SZ in Figure \ref{fig:wo-shape-no-freq}.  
This case does not successfully recover kSZ.  
There is a systematic bias in the SZ, but 
statistical errors dominate.  The bins are strongly correlated.  
Systematic biases dominate statistical errors in the recovery of the 
point source power spectrum.  The relative bias is greatest at 145 
GHz, where the point source contribution is the least, and decreases 
at higher frequencies. Note that in case 4 especially, the Fisher matrix 
characterizes the errors poorly.

\begin{figure*}
\begin{center}
\includegraphics[width=\widefig]{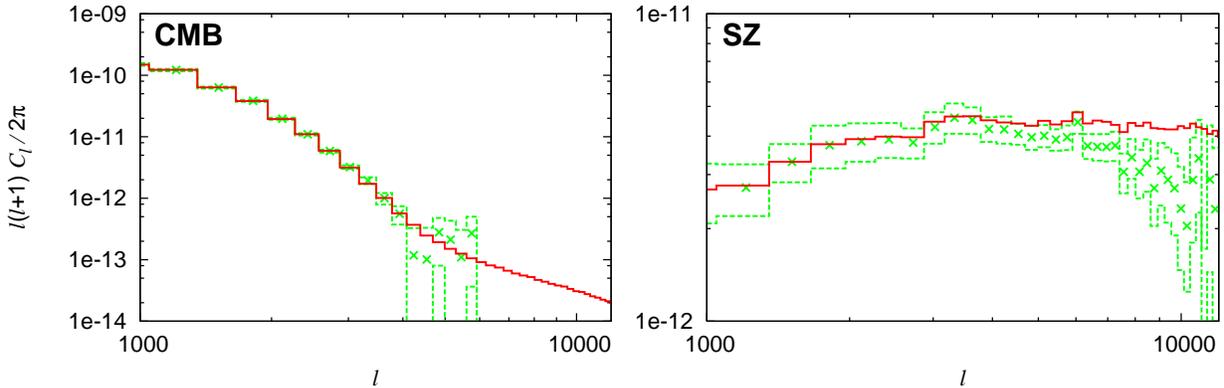}\\
\end{center}
\caption{
Estimates of the power spectra in case 6, where we assume the shape 
of the kSZ and point source power spectra, but assume nothing about the
point source frequency dependence. 
\label{fig:estpower-case6}
}
\end{figure*}

In cases 5 and 6, the kSZ is separated from the CMB.  This is 
necessary to allow  parameter estimation based on the CMB.  Because 
the CMB and kSZ are degenerate in frequency, this requires a model for 
the kSZ power spectrum.  The estimator in this case uses the high $l$ 
portion of the kSZ spectrum, where CMB is not significant, to 
calibrate the model of kSZ.  kSZ is then subtracted at low $l$.  We 
use a 1-parameter model for the kSZ power spectrum where the shape is 
known but the amplitude is not.  The approach generalizes to a 
multiparameter model.  We do not plot case 5.  For case 6, we plot 
the recovered power spectrum (Figure \ref{fig:estpower-case6}).  We do not 
detect a bias on the CMB in either case, although the SZ power 
spectrum is biased in case 6 as it was in case 3.  The CMB power
bins for $l>3000$ are correlated by the kSZ template subtraction.  SZ bins 
are also strongly correlated.


We now summarize our power spectrum analysis results.  Given a good 
understanding of the frequency dependence of point sources, we may 
successfully estimate the power spectrum of each component, provided 
we treat the CMB and kSZ as a single component.  
With no prior knowledge of the frequency dependence of point sources, 
we may make a somewhat biased estimate of the power spectra if we 
know the shape of the point source power spectrum.  Knowing neither
the frequency dependence nor the shape of sources, 
we detect SZ, but cannot positively identify kSZ. For us 
to have confidence in our estimates, we must trust our knowledge of 
the point source frequency dependence.  This prospect is unclear.  If 
we do not have confidence in our frequency knowledge, we may still 
make an estimate of the power spectrum, but it will be somewhat 
biased.
In our more optimistic cases, if we want to estimate CMB without kSZ, 
we require a parameterized model for kSZ.  The power spectrum 
template will likely come from simulation.  However, any 
parameterized fit to the power spectrum of the signal components may 
introduce a bias. 

\subsection{Measuring primordial amplitude and slope} 
\label{subsec:cosmopar}

In our model,  the kSZ power is already comparable to the size of the 
CMB error bars at $l \sim 2000$, so if we do not know 
the kSZ power spectrum at all then we should not use the CMB power 
spectrum 
beyond there.  With a model for the kSZ power spectrum we may use bins 
with higher $l$, as in our cases 5 and 6.  
To illustrate this difference, we compute the 
covariance matrix for primordial amplitude $A$ and slope $n$ about 
the fiducial values $A=1$ ({\sl COBE} normalized) and $n=1$.  To 
convert from errors on binned power to errors on parameters we use
\be
{\sf C}^{-1}_{qq'} = \sum_{ll'} \frac{dC_l}{dq} {\sf C}^{-1}_{ll'} 
\frac{dC_l}{dq'}
\ee
where ${\sf C}_{qq'}$ is the covariance matrix for parameters $q$,  
${\sf C}_{ll'}$ is the covariance matrix for the binned power, 
consisting of the covariances computed for CMB.  For this we use the 
CMB block of the covariance matrix for case 6. 
 (The result using case 5 is similar.) We compute the derivatives of 
the power spectrum using CMBFAST.  Then for primordial amplitude and 
slope, we have the following covariance matrix if we do not extract 
kSZ.
\be
{\bf \sf C}(l<2000) = \left\{
\begin{array}{c|cc}
    &    A   &   n  \\ \hline
A   &    2 \times 10^{-2}   &   -4 \times 10^{-3}  \\
n   &                       &  1 \times 10^{-3}  \\
\end{array}\right\}.
\ee
However, if we can extract the CMB, the covariance improves.
\be
{\bf \sf C} = \left\{
\begin{array}{c|cc}
    &    A   &   n  \\ \hline
A   &    4 \times 10^{-3}   &   -9 \times 10^{-4}  \\
n   &                         &  2 \times 10^{-4}  \\
\end{array}\right\} .
\ee
Thus the covariance increases by roughly a factor of 5 for $A$ and 
$n$ if the kSZ is not extracted. If inhomogeneous reionization raises 
the kSZ power, shortening further the lever arm of the CMB, it 
becomes more important to extract kSZ.

To gain this additional information from the CMB on scales $l > 2000$ 
we need a template for kSZ.  
This template will come from simulations, and may be calibrated on 
the kSZ 
measured at higher $l$.  However, there will be uncertainties in the 
template
because of missing physics, such as patchy reionization. 
Therefore the power spectrum analysis should be checked
in as many ways as possible against the data.  
One way we explore in the next section is using the 
non-Gaussian information.

\section{One-point distribution function analysis} \label{sec:ng-pdf}

CMB and kSZ cannot be distinguished in a multifrequency analysis.  We 
have seen this in our power spectrum analysis in the previous 
section, in which we could only estimate CMB separately from kSZ by 
knowing the shape of the kSZ power spectrum beforehand.  However, if 
the data do not represent a Gaussian random field, and our data do 
not, then the power spectrum does not extract all the information 
available.

Statistics which highlight non-Gaussianity are also particularly 
interesting because in the standard picture, primary CMB fluctuations 
are Gaussian.  Any non-Gaussianities indicate deviations from the 
standard picture for primary fluctuations, or in our case, indicate 
the presence of secondary sources of anisotropy.

In principle then, it is possible to use non-Gaussianities to tease 
apart the CMB and kSZ.  In the following, we show this is possible, 
and explore one method for doing so.  The statistic we choose is the 
histogram of pixel temperatures, that is, the 1-point 
probability distribution 
function (pdf).  In the context of weak lensing, the pdf provides 
stronger constraints than simple statistics such as skewness 
\cite{2000ApJ...530..547J}.  This makes sense because the pdf 
incorporates all the moments of the field evaluated at a single 
point, such as variance, kurtosis, and so on.  For a Gaussian field 
the pdf is Gaussian.  For a highly non-Gaussian field, such as kSZ, 
the pdf function may show strong tails.  The information in 
the pdf is complementary to the information in the power spectrum.  
The pdf contains information about non-Gaussianities, but not about 
spatial correlations, because it is evaluated at a single point. The 
power spectrum contains information about spatial correlations, but 
not about non-Gaussianities.
 
\begin{figure} 
\begin{center}
  \includegraphics[width=\narrowfig]{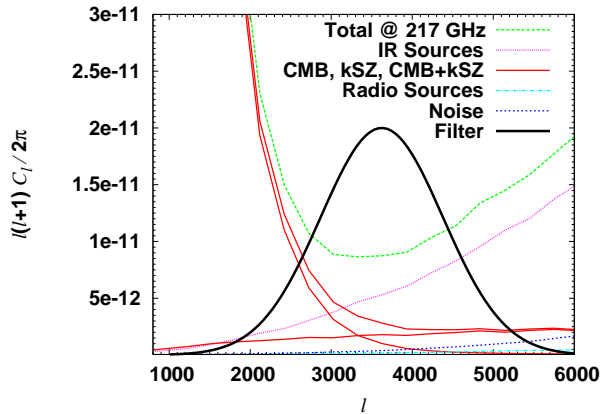}
\end{center}
  \caption{The spatial filter used for the non-Gaussian analysis and the 
components at 217 GHz.  The vertical scale of the filter in 
this plot is arbitrary.   
}\label{fig:window}
\end{figure}

Below, we devise a method for examining the pdf of our maps to 
extract parameters of the component signals, given a set of 
templates.  Then we apply this method.  We want to separate CMB from 
kSZ, so we consider the sky at 217 GHz.  This is simplest because it 
eliminates the signal from thermal SZ.  We want to focus on scales 
where neither the CMB nor kSZ completely dominates the other.  Thus, 
before evaluating the pdf, we spatially filter the maps.  In Fourier 
space, we multiply the maps by a Gaussian centered at $l \approx 
3600$ with standard deviation $\sigma_l \approx 750$.  In figure 
\ref{fig:window}, this filter is shown with the the power spectra of 
the various components of the sky at 217 GHz.

In the next sections, we describe the pdfs of the major signals at 
217 GHz in this window, describe our method for separating kSZ from 
CMB, and apply it to our simulated maps.

We make no claim that this is an optimum method for separating CMB 
and kSZ.  At this point, this method is intended as a demonstration 
of the power of using simple statistics which are sensitive to 
non-Gaussianities.

\subsection{Signal one-point distribution functions}

\begin{figure*} 
\begin{center}
  \includegraphics[width=\widefig]{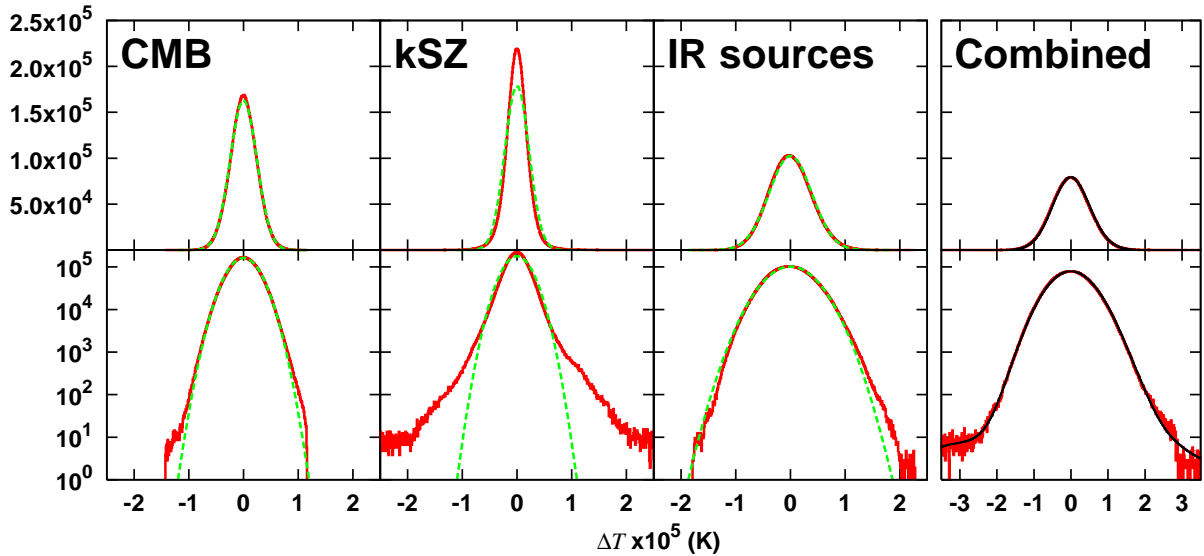}
\end{center}
\caption{The histograms of the major signals at 217 GHz, after 
spatial filtering.  Each pair of plots shows the same data, except 
the vertical axis above has a linear scale and below has a log scale. 
 The distributions have been normalized to unity.  The CMB is well 
fit by a Gaussian.  kSZ is poorly fit by a Gaussian with the same 
variance.  Infrared point sources are slightly skewed in the positive 
direction, compared to a Gaussian.  At right we show the histogram of 
the sum of these maps, along with our best fit pdf from fitting and 
convolving the template pdfs, as described in the 
text.}\label{fig:sigpdf}
\end{figure*}

Here we examine the pdfs of the signals with the goal of
identifying features that distinguish CMB and kSZ.  In 
addition, the infrared sources will be an unavoidable 
contaminant at 217 GHz.  The radio sources are not very significant, 
and as we have mentioned, may be helped by additional masking.  The 
noise after spatial filtering is very low.  After spatial filtering, 
we show the histograms for the major signals at 217 GHz in Figure 
\ref{fig:sigpdf}.

The pdf of the unlensed CMB is Gaussian.  
After lensing, the distribution is still Gaussian because lensing 
simply remaps locations on the sky.  After spatial filtering, the 
lensed CMB need not have a Gaussian distribution function, but we 
find the deviations in this case to be insignificant.

The distribution of kSZ should be symmetric, because there should be 
roughly as many clusters moving towards us as away.  We expect the 
pdf of kSZ, or any signal sourced by discrete objects, to have a 
non-Gaussian point distribution function.  This is because the signal 
strength is highly localized: the values of pixels are extreme in the 
direction of the object, and zero elsewhere.  The kSZ histogram shows 
tails well beyond that expected from a Gaussian pdf of the same 
variance.  A pdf with power law tails, such as Student's 
$t$-distribution, is a much better fit, although we find the 
$t$-distribution to be too poor of a model to apply our method.

We might expect the point sources to have a non-Gaussian pdf, since
 they are positive definite, which would seem 
to argue for strong skewness.  This is counteracted by two effects.  
For a high density of sources, several sources are 
included in each beam, and the non-Gaussianity is diminished due to 
central limit theorem considerations.  In addition, our spatial 
filtering removes long wavelength modes, causing ringing about each 
source which can wash out skewness.  We show the histogram for 
infrared point sources, after spatial filtering, in figure 
\ref{fig:sigpdf}; it is nearly Gaussian.   A wider 
window function reduces ringing, and produces a point source pdf 
which is more positively skewed.

We see that the tails of the kSZ distribution are a unique feature with 
which we can try to distinguish it from the other signals.

\subsection{Likelihood}
  
Given a set of pixelized data, we want to examine the likelihood of a 
trial distribution function.  Let $P_\mathbf{a}(\Delta T)$ be a trial 
pdf,  which depends on a set of parameters $\mathbf{a}$.  For now, think of 
it as 
an abstract parameterized model for our data histogram.  We construct our 
trial pdf in the next section.  For a set of 
independent pixels $\Delta T_i$, the likelihood of parameters 
$\mathbf{a}$ is given by 
  \be
  L(\mathbf{a}) = \prod_i P_\mathbf{a}(\Delta T_i),
  \ee
which is just the probability of the first pixel, times that of the 
second, and so on.  Computationally it is more convenient to work 
with the logarithm of the likelihood.  We also work with a binned 
pdf, so we combine all the pixels for each bin, and sum over bins:
\be
\log L(\mathbf{a}) = \sum_j N_j \log P_\mathbf{a}(\Delta T_j), 
\label{eqn:loglike}
\ee
where $N_j$ is the number of pixels in histogram bin $j$, and $\Delta 
T_j$ is the value at the center of bin $j$.  This gives us the 
likelihood of parameters $\mathbf{a}$ for model pdf $ 
P_\mathbf{a}(\Delta T)$ in terms of the histogram of pixel values 
$N_j$.

If the pixels are not all independent, this is not the correct 
likelihood function.  The true likelihood would take into account the 
covariance between pixels (or alternatively, between bins).  However, 
we find in practice that the peak of the given likelihood function in 
(\ref{eqn:loglike}) is a good enough estimate for this application.  
The correlation length of our maps after spatial 
filtering is much less than the size of the maps. 
Thus our maps contain a large number of uncorrelated patches.  
If these patches are thought of as larger, independent pixels, then 
the likelihood we have written has the right peak, but may have a wrong 
normalization and width.  In our method, we use only the location of the 
likelihood peak, while the errors are assessed using Monte Carlo simulations.

Our goal is to find the set of parameters $\mathbf{a}$ which maximize 
the likelihood.  We use Powell's direction set method 
\cite{1986nras.book.....P}.  Because the data (the histogram of pixel 
temperatures) is noisy, the likelihood function is plagued by local 
maxima.  We repeat our maximum likelihood search with randomized 
initial positions and search directions in parameter space, to ensure 
that the maximum we find is reliable.

We compute errors on our parameters by the bootstrap method.  We 
sample our sky map realizations (with replacement) to generate  
synthetic data histograms.  We find the maximum likelihood parameters 
for each of these synthetic histograms, and use the distribution of 
these to describe the probability distribution for the parameters.

\subsection{Separation by distribution function}

To evaluated the likelihood function (\ref{eqn:loglike}), we need a 
parameterized model pdf for the data histogram at 217 GHz.  To get 
unbiased estimates of the parameters, our model must faithfully 
reproduce the data.  We proceed by making models of each signal in 
turn.  The pdf of the sum of independent random variables is given by 
the convolution of the pdfs of the individual random variables. 
Therefore, by modeling the pdfs of the individual signals, we can 
produce a trial pdf for the total data by convolution.  We use binned 
functions for our models, and evaluate the convolution with fast 
Fourier transforms.  Our procedure is not very sensitive to the width 
of the bins unless the model pdfs contain sharp features.  In this 
application, our pdfs are smooth functions, and we find that roughly 
one thousand bins covering the pdf suffices.

We make a 2 parameter model for the pdf.  Our parameters are the 
quantities we seek to estimate, in this case the standard deviation 
of the CMB, which we label $\sigma_\textrm{\scriptsize CMB}$, and the 
standard deviation of the kSZ, which we label 
$\sigma_\textrm{\scriptsize kSZ}$.  For the model pdf of the CMB we 
simply use a Gaussian with variance $\sigma_\textrm{\scriptsize 
CMB}^2$.  For the kSZ pdf, we use the actual histogram of the kSZ as 
a template.  In the real experiment, this would not be available, but 
could be based on simulations.  To generate a distribution with a 
given standard deviation we rescale the template, preserving the 
distribution's unit integral.   For the model pdf of the point 
sources, we again use the actual point source histogram as a 
template.  If we understand the point source population well enough to 
make a good power spectrum analysis, such that we wish to attempt a 
separation of CMB and kSZ, the we should be able to model the point 
source pdf.  Note that our non-Gaussian analysis will provide a 
constraint on $\sigma_\textrm{\scriptsize CMB}$ and 
$\sigma_\textrm{\scriptsize kSZ}$ individually.  Contrast this to the 
power spectrum analysis, which constrains only the total variance $ 
\sigma_\textrm{\scriptsize CMB}^2 + \sigma_{\rm kSZ}^2$, albeit in a 
narrower window of multipoles.

\begin{figure} 
\begin{center}
  \includegraphics[width=\narrowfig]{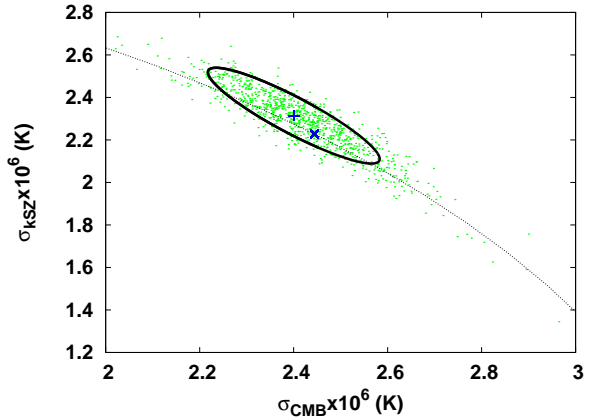}
\end{center}
\caption{Separating CMB from kSZ using the pdf.  This plot shows the 
result of our fitting procedure for the standard deviation of the kSZ 
and CMB.  The ``$+$'' shows the best fit point.  The ``$\times$'' 
marks the true value.  The dots show the distribution of the fits for 
1000 bootstrap realizations.  The contour denotes the 68\% region 
computed from the covariance of the bootstrap points ($\chi^2$ with 2 
degrees of freedom). The fine dotted line shows the circle of constant
total variance which passes through the true value.}\label{fig:ng-fit}
\end{figure}

We apply our maximum likelihood parameter estimate to this model.  
The result is a $>10\sigma$ detection of the CMB and a $>7\sigma$ 
detection of the kSZ in our window (see Figure \ref{fig:ng-fit}).  
The estimate is somewhat degenerate in the direction which preserves 
the total variance $ \sigma_\textrm{\scriptsize CMB}^2 + \sigma_{\rm 
kSZ}^2$.

The power of this method arises from the template for kSZ, which 
fixes the relation between the kSZ pdf's variance and tails.  If this 
relationship is not fixed, then a narrow kSZ distribution with wide 
tails could mimic a wide kSZ distribution with narrow tails, by 
adjusting the variance of the CMB pdf to compensate.  Also, it is 
crucial to have the point source contribution constrained.  Otherwise 
it would be degenerate with the CMB.

We must depend on simulations for the relationship between the 
variance and tails of kSZ.  This non-Gaussian method may also be 
viewed as a consistency check for kSZ template simulations.  A 
simulation of kSZ must both satisfy the power spectrum at high $l$, 
and provide the correct non-Gaussianity where kSZ and CMB 
overlap.  If the measurements differ, one should not have confidence 
in the template power spectrum for kSZ provided by the simulation.

\section{Conclusions} \label{sec:conclusions}
We have developed simulations of the ACT experiment
in three frequency bands and analyzed the ability of ACT
to extract the power spectrum and non-Gaussian statistics from it. 
The simulations include 
CMB, kSZ, SZ, and radio and infrared point sources.  Secondary 
anisotropies dominate the CMB over a wide range of interest.  These 
in turn are heavily polluted by point sources.  Infrared point 
sources must be dealt with in the power spectrum, and cannot be 
completely dealt with by masking.  
While we included the latest constraints on the secondary anisotropies, 
significant uncertainties on the amplitude and correlation 
properties remain and could affect the conclusions reached in this paper. 
Particularly uncertain is the contribution of patchy reionization to kSZ.

An optimal multifrequency filter can extract the power spectrum, but 
the results are subject to assumptions put into the analysis.  
We considered several cases, making different assumptions about our 
knowledge of contaminating extragalactic point sources. For optimal 
extraction knowing both the 
mean and scatter of the frequency dependence of point sources is 
crucial.  If we do not use a prior model for the frequency dependence
of the point sources 
we may still extract the power spectrum assuming spatial correlations
are known, but the results may still be biased. Furthermore,  
using only frequency information in a power spectrum analysis, kSZ 
and CMB cannot be separated using data alone.  This degrades the 
ability to measure the CMB at $l>2000$ and to determine parameters from 
it.

We also explored the ability to use the non-Gaussian information 
to constrain further the individual components. 
Non-Gaussianities in the histogram of pixel temperatures at 217 GHz 
should be observable by ACT.  Non-Gaussianities created by kSZ can allow
one to distinguish it from Gaussian CMB on scales where the two cannot be 
separated with the power spectrum analysis, assuming that simulations 
are able to provide an accurate template of kSZ.  
It is unclear how well a kSZ template from simulations will work
when applied to, for example, kSZ created by patchy reionization.
Therefore, our results are preliminary and more 
investigation of these techniques is needed.
Combining the power spectrum template analysis 
with the non-Gaussian template analysis
allows one to perform consistency checks among these methods. 

Our results suggest that extracting the primordial power spectrum 
information at high precision 
from small scale primary CMB will be challenging. 
Scatter in the frequency scaling of point sources, as well as their 
possible correlations, make the point source separation from CMB difficult.  
Assuming this is accomplished, removing kSZ is even more difficult and 
can only be done with reliable templates for the power spectrum and for 
non-Gaussian signals. The final precision is likely to be limited by 
the modeling uncertainties and not by the statistical precision of the 
observations. It is hard to prognosticate the final accuracy given 
how uncertain the kSZ amplitude from patchy reionization is. 
In this context it is worth mentioning that CMB polarization 
has fewer secondary anisotropies to worry about than the CMB temperature 
anisotropy itself and so may ultimately be 
more promising as a tool to study the primordial power spectrum 
fluctuations on small scales. 

\section*{Acknowledgments}
We are grateful to Pengjie Zhang, Mike Bichan and Ue-Li Pen for 
assistance with 
the hydrodynamical simulations, to Tomonori Totani, for providing 
source 
counts for infrared sources, and to Lyman Page, for encouragement at 
the outset.  
KMH wishes to thank Nihkil Padmanabhan, Christopher Hirata, and 
Mustapha Ishak for useful conversations.  
For most of this work, KMH was supported by an NSF graduate research 
fellowship.
US is supported by the Packard Foundation,
NASA NAG5-1993,  NASA NAG5-11489, and NSF CAREER-0132953.

\bibliographystyle{unsrt}
\bibliography{cosmo,cosmo_preprints,reference}

\end{document}